\pdfoutput=1

\documentclass[journal,twoside]{IEEEtran}

\usepackage{flushend} 
\usepackage{ifpdf}
\usepackage{cite}
\ifCLASSINFOpdf
 	\usepackage[pdftex]{graphicx}
 	\DeclareGraphicsExtensions{.pdf,.jpeg,.png}
\else
 	\usepackage[dvips]{graphicx}
 	\DeclareGraphicsExtensions{.eps}
\fi
\graphicspath{{./figures/}}
\usepackage[cmex10]{amsmath}
\usepackage{amsfonts} 
\usepackage{amssymb} 
\usepackage{amsthm} 
\usepackage{array}
\usepackage{arydshln}
\usepackage{url}

\newcommand{\vect}[1]{\mathbf{#1}}
\newcommand{\matr}[1]{\mathbf{#1}}
\newcommand{\set}[1]{\mathcal{#1}}
\newcommand{\code}[1]{\mathcal{#1}}

\newcommand{\Z}{\mathbb{Z}}
\newcommand{\F}{\mathbb{F}}

\newcommand{\stacktwo}[2]{\genfrac{}{}{0pt}{}{#1}{#2}}
\newcommand{\minstarnoarg}{\operatorname{min}^{*}}
\newcommand{\minstar}[1]{\underset{#1}{\operatorname{min}^{*}}}
\newcommand{\wH}{w_{\mathrm{H}}}
\DeclareMathOperator{\wt}{wt}
\DeclareMathOperator{\adj}{adj}
\DeclareMathOperator{\perm}{perm}
\DeclareMathOperator{\sign}{sign}

\theoremstyle{plain}
\newtheorem{thm}{Theorem} 
\newtheorem{lem}[thm]{Lemma}

\theoremstyle{definition}
\newtheorem{exmp}{Example} 

\theoremstyle{remark}
\newtheorem{remk}{Remark} 

\makeatletter
\newlength \figwidth
\if@twocolumn
\else
\fi
\makeatother

\newcommand{\mysmallarraydecl}{\renewcommand{%
\IEEEeqnarraymathstyle}{\scriptstyle}%
\renewcommand{\IEEEeqnarraytextstyle}{\scriptsize}%
\renewcommand{\baselinestretch}{1}%
\settowidth{\normalbaselineskip}{\scriptsize
\hspace{\baselinestretch\baselineskip}}%
\setlength{\baselineskip}{\normalbaselineskip}%
\setlength{\jot}{0.25\normalbaselineskip}%
\setlength{\arraycolsep}{4pt}}

\usepackage{hyperref}
\hypersetup{
  pdfinfo={
    Title={Bounds on the Minimum Distance of Punctured Quasi-Cyclic LDPC Codes},
    Author={Brian K. Butler},
    Subject={Information Theory, Error Correction Coding},
    Keywords={LDPC; Error Correction; Coding; AR4JA; Protograph}
  }
}

\begin{document}
\title{Bounds on the Minimum Distance of Punctured Quasi-Cyclic LDPC Codes}

\author{Brian~K.~Butler,~\IEEEmembership{Senior~Member,~IEEE,}
Paul~H.~Siegel,~\IEEEmembership{Fellow,~IEEE}
\thanks{This work was presented in part at the IEEE International Symposium on Information Theory, Austin, Texas, June 2010.}
\thanks{The authors are with the Department of Electrical and Computer Engineering,
University of California, San Diego (UCSD), La Jolla, CA 92093 USA (e-mail: butler@ieee.org, psiegel@ucsd.edu).}
\thanks{This work was supported in part by the Center for Magnetic Recording Research at UCSD and by the National Science Foundation (NSF) under Grant CCF-0829865 and Grant CCF-1116739.}
}

\maketitle

\ifCLASSOPTIONpeerreview
	\markboth{Bounds on the Minimum Distance of Punctured Quasi-Cyclic LDPC Codes\quad Version:  February 19, 2013}%
	{Bounds on the Minimum Distance of Punctured Quasi-Cyclic LDPC Codes}
\else
	\markboth{\MakeLowercase{updated submission to} \textit{IEEE Transactions on Information Theory} \;\;\;\;Version:  February 19, 2013}%
	{Butler and Siegel: Bounds on the Minimum Distance of Punctured Quasi-Cyclic LDPC Codes}
\fi

%

\begin{abstract}
Recent work by Divsalar et al.\ has shown that properly designed protograph-based low-density parity-check (LDPC) codes typically have minimum (Hamming) distance linearly increasing with block length. 
This fact rests on ensemble arguments over all possible expansions of the base protograph. 
However, when implementation complexity is considered, the expansions are frequently selected from a smaller class of structured expansions.
For example, protograph expansion by cyclically shifting connections generates a quasi-cyclic (QC) code. 
Other recent work by Smarandache and Vontobel has provided upper bounds on the minimum distance of QC codes. 
In this paper, we generalize these bounds to punctured QC codes and then show how to tighten these for certain
classes of codes. 
We then evaluate these upper bounds for the family of protograph codes known as AR4JA codes
that have been recommended for use in deep space communications in a standard established 
by the Consultative Committee for Space Data Systems (CCSDS).  
At block lengths larger than 4400 bits, these upper bounds fall well below the ensemble lower bounds.
\end{abstract}


%
\IEEEpeerreviewmaketitle

\section{Introduction}
\label{sect1}
\IEEEPARstart{L}{ow-density} parity-check (LDPC) codes originated in
the seminal work by Gallager \cite{Gal63} over 50 years ago. 
The study of these codes remained largely dormant for decades, with the important exception of Tanner's work on graph-based code constructions \cite{Tan81}.
At low SNR, properly designed LDPC codes exhibit good performance with practical, iterative message-passing decoders.
However, at higher SNRs, they may suffer from an abrupt change in the slope of the error-rate curve, a phenomenon known 
as an error floor. The floor can be attributed, in part, to the existence of certain properties of the Tanner graph that is associated with
a chosen parity-check matrix and upon which the decoder operates. 
Techniques that reduce the occurrence of short cycles in the Tanner graph, for example \cite{RichFloors,PEG}, 
have been shown to mitigate the error floor phenomenon.
Specifically, the ACE algorithm \cite{ACE} for placing edges in a graph-based code brings down the error floor substantially by 
preventing short cycles from clustering around low-degree variable nodes.

Another code property that limits error performance at high SNR is the minimum (Hamming) distance between codewords. The minimum distance is also important in understanding the likelihood of undetected errors, a critical concern in many applications. 
Yet, relatively little attention has been paid to analyzing the minimum distance of LDPC codes and to developing LDPC code design methodologies that ensure large minimum distance. 
MacKay and Davey introduced upper bounds on the minimum distance for a class of codes that included quasi-cyclic (QC) LDPC codes in \cite{MKseagate}.
Notable later work appears in \cite{Tan04,FossQC}.
Of particular relevance to this work are the upper bounds of Smarandache and Vontobel \cite{SmaVon12} which allow for more variation in the underlying protograph used to represent the code.

Another line of research has shown that most codes in the ensemble of protograph-based codes characterized by a limited number of degree-two variable nodes have minimum distance that increases linearly with block length \cite{DivDol06,DivDol09,AbuDiv11}.
The family of LDPC codes known as AR4JA codes, recommended for deep-space communications by the Consultative Committee for Space Data Systems (CCSDS) \cite{CCSDS11}, are obtained by puncturing QC-LPDC codes designed from protographs in this ensemble.
The selected protographs are expanded to the actual codes (in two stages) using the ACE algorithm to place the edges with QC constraints.

In this paper, we extend the bounds of \cite{SmaVon12} to the general class of punctured QC-LDPC codes, and show that these bounds can be tightened in cases where the protomatrices associated with the underlying protograph contain many zero entries. 
Much of our methodology parallels \cite{SmaVon12}, with our extensions motivated by an interest in bounding the actual minimum distance of the AR4JA codes specified in the CCSDS standard. 
Somewhat surprisingly, the application of our methodology to the protomatrices underlying the AR4JA constructions for code rates $1/2$, $2/3$, and $4/5$ yields upper bounds of $66$, $58$, and $56$, respectively, independent of the code block length. 
For large block lengths, these bounds fall well short of the linearly-growing ensemble lower bound mentioned above. 
Finally, using slight modifications of previously proposed search techniques, we identify specific codewords for each of these code rates at two of the standardized code lengths. The weights of these codewords validate the upper bounds and suggest that the upper bounds may be fairly tight.

The remainder of the paper is organized as follows. Section~\ref{sect2} provides a review of protograph-based LDPC code design, as well as the specific family of AR4JA codes. Section~\ref{sect3} provides the necessary mathematical background on the polynomial representation and properties of QC-LDPC codes obtained by the expansion of protographs in which the expansion is based on circulant matrices. In Section~\ref{sect4}, we review the upper bounds on the minimum-distance of QC-LDPC codes in \cite{SmaVon12}, and then develop the necessary algebraic results to generalize these bounds to punctured QC-LPDC codes. 
Section~\ref{sect5} describes techniques that can produce tighter upper bounds for protographs with specific properties, including some of the AR4JA protographs. 
In Section~\ref{sect6}, we apply our methods to calculate upper bounds on the minimum distance of the codes in the CCSDS standard, which are obtained by a two-step QC expansion of the AR4JA protographs. 
In Section~\ref{sect_search}, we use computer search to find low-weight codewords for several AR4JA codes, compare them to our length-independent bounds, and reconcile them with the ensemble minimum-distance lower bounds \cite{DivDol06,DivDol09,AbuDiv11}. 
We also examine the girth of AR4JA codes. Section~\ref{sect_concl} concludes the paper. 
\begin{figure}[!t]
\centering{
{\includegraphics[width=1.4in]{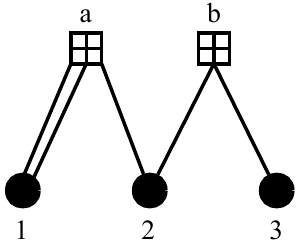}} 
\quad
{\raisebox{0.6 in}
{$\matr{A}= \begin{bmatrix} 2&1&0\\
0&1&1 \end{bmatrix}$}%
}
}
\caption{Simple protograph $G$ and corresponding protomatrix $\matr{A}$.}
\label{fig_proto1}
\end{figure}

\section{Protographs and AR4JA}
\label{sect2}
Protographs were introduced as a way to impart structure to the interconnectivity of graph-based codes \cite{Tho03}. 
Protographs themselves are a subset of multi-edge type graphs \cite[ch.~7]{RichUrb}.

A \emph{protograph} is essentially a Tanner graph with a relatively small number of nodes. More specifically, 
a protograph, $G = \left(V, C, E\right)$, consists of a set of variable nodes $V$,
a set of check nodes $C$, and a collection of edges $E$.
Each edge, $e \in E$, connects a variable node, $v_e \in V$, to a check node, $c_e \in C$. 
Protographs have the additional property that parallel edges are permitted. Moreover, variable nodes in $V$
can be designated as punctured; \textit{i.e.}, the corresponding bits are not included in the transmitted codeword.

A simple protograph $G$ is shown in Fig.~\ref{fig_proto1} with three variable nodes, two check nodes, and five edges. 
The accompanying \emph{protomatrix} $\matr{A}$ fully describes the protograph structure.
The entry in the $j$th row and $i$th column of the protomatrix $\matr{A}$ indicates the number of edges connecting 
the $j$th check node to the $i$th variable node within the corresponding protograph.

A \emph{derived graph} is constructed by replicating the protograph a specified number of times and interconnecting the copies of the variable and check nodes in a manner consistent with the topology of $G$. 
In this example, all copies of check node $a$ are called ``type $a$'' check nodes. 
Similarly, all copies of variable node $1$ are called ``type $1$'' variable nodes.
The replicas of an edge connecting check node $a$ and variable node $1$ form a so-called edge set, and their connected check nodes may be permuted within the set of ``type $a$'' check nodes. 
(The term ``type,'' when used in Section~\ref{sect6} to classify matrices as in \cite{SmaVon12}, is unrelated.)
Application of this interconnection procedure to all replicas ensures that node degrees and connectivity by node types of the original protograph are maintained in the resulting derived graph. 
The corresponding linear code is referred to as a \emph{protograph code}. 

\begin{figure}[!t]
\centering
\includegraphics[width=1.81in]{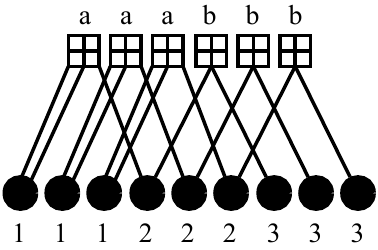} 
\caption{Protograph $G$ replicated $N=3$ times.}
\label{fig_proto2}
\end{figure}

\begin{figure}[!t]
\centering
\includegraphics[width=1.81in]{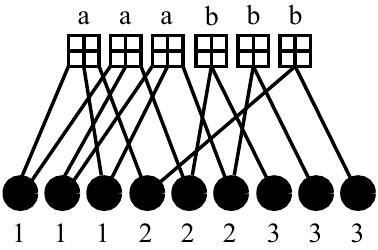} 
\caption{Derived graph obtained by edge set permutations that preserve degree and interconnectivity among node types.}
\label{fig_proto3}
\end{figure}

Figs.~\ref{fig_proto2} and \ref{fig_proto3} illustrate the process of making $N=3$ copies of the protograph of Fig.~\ref{fig_proto1} 
and interconnecting them to generate the derived graph. 
The parity-check matrix corresponding to the derived graph of Fig.~\ref{fig_proto3} is shown below, divided into 
submatrices so the relationship to the protomatrix $\matr{A}$ of Fig.~\ref{fig_proto1} is evident:
\begin{equation*}
\label{H69}
    \matr{H}=
         \left[
           \begin{array}{ccc;{2pt/2pt}ccc;{2pt/2pt}ccc}
             1 & 0 & 1 &  1 & 0 & 0 &  0 & 0 & 0 \\
             1 & 1 & 0 &  0 & 1 & 0 &  0 & 0 & 0 \\
             0 & 1 & 1 &  0 & 0 & 1 &  0 & 0 & 0 \\
          \hdashline[2pt/2pt]
             0 & 0 & 0 &  0 & 1 & 0 &  1 & 0 & 0 \\
             0 & 0 & 0 &  0 & 0 & 1 &  0 & 1 & 0 \\
             0 & 0 & 0 &  1 & 0 & 0 &  0 & 0 & 1
           \end{array}
         \right].
\end{equation*}

Protograph-based code design allows for the introduction of degree-one variable nodes and punctured variable nodes in a structured way. 
With regard to degree-one nodes, recall that the optimization of irregular LDPC codes by density evolution typically avoids degree-one variable nodes,
since they impart an error rate floor on randomly constructed codes even as block length grows toward infinity \cite[p.~161]{RichUrb}.
However, density evolution often produces a significant fraction of degree-two variable nodes. 
This suggests that the incorporation of degree-one variable nodes may offer a potential benefit in code performance, as noted in \cite[p.~382]{RichUrb}. Another advantage of protographs is that the corresponding iterative decoder implementation may be less complex than that of ``random'' LDPC codes because of the structure imposed on the node interconnections.

\begin{figure}[t]
\centering
\includegraphics[width=0.8\figwidth]{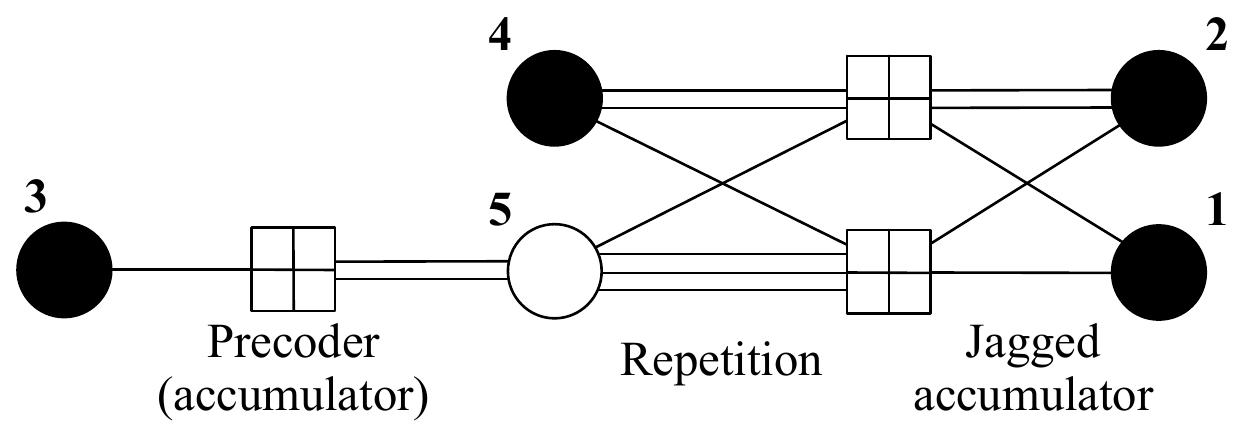} 
\caption{AR4JA protograph, rate-$1/2$.}
\label{fig_ar4ja12}
\end{figure}

\begin{figure}[t]
\centering
\includegraphics[width=0.8\figwidth]{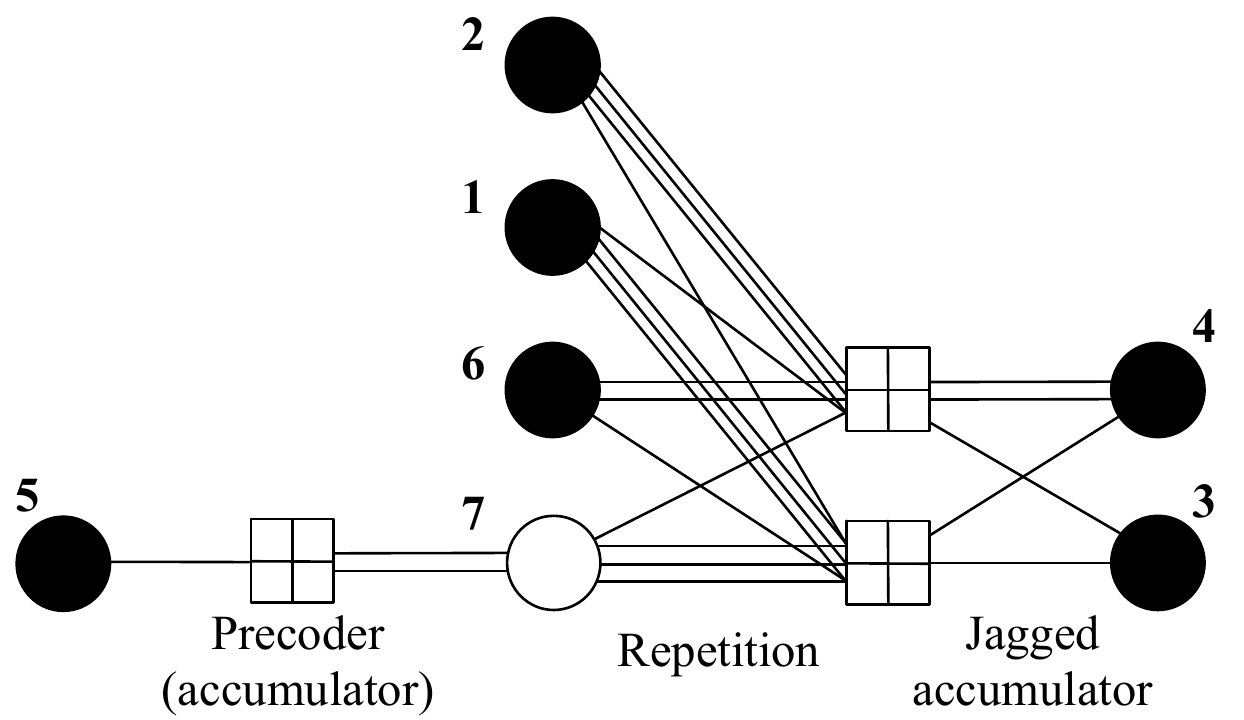} 
\caption{AR4JA protograph, rate-$2/3$.}
\label{fig_ar4ja23}
\end{figure}

The protograph for the rate $r=1/2$ AR4JA code \cite{DivDol06} is shown in Fig.~\ref{fig_ar4ja12}.
We follow the convention of representing the transmitted variable nodes as solid circles and the punctured variable nodes as unfilled circles. 
The protograph of the rate-$1/2$ code is extended to rate-$2/3$ by adding two degree-four variable nodes as shown in Fig.~\ref{fig_ar4ja23}. 
The corresponding protomatrices are  
\begin{equation}
\label{proto_ar4ja12}
\matr{A}_{r=1/2}=\begin{bmatrix}
0&0&1&0&2\\
1&1&0&1&3\\
1&2&0&2&1\end{bmatrix}
\end{equation}
and
\begin{equation}
\label{proto_ar4ja23}
\matr{A}_{r=2/3}=\begin{bmatrix}
0&0&0&0&1&0&2\\
3&1&1&1&0&1&3\\
1&3&1&2&0&2&1\end{bmatrix},
\end{equation}
respectively.
The numerical labels of the variable nodes in the figures correspond to the columns in the protomatrices, enumerated from left to right, 
and the rows of the protomatrix correspond to the check nodes.

The AR4JA family of protographs further extends the code rate options by adding an additional four degree-$4$ variable nodes.
The rate-$4/5$ protograph has $11$ variable nodes altogether, as can be seen from its protomatrix,
\begin{equation}
\label{proto_ar4ja45}
\matr{A}_{r=4/5}=\begin{bmatrix}
0\ \ 0\ \ 0\ \ 0\ \ 0\ \ 0\ \ 0\ \ 0\ \ 1\ \ 0\ \ 2\\
3\ \ 1\ \ 3\ \ 1\ \ 3\ \ 1\ \ 1\ \ 1\ \ 0\ \ 1\ \ 3\\
1\ \ 3\ \ 1\ \ 3\ \ 1\ \ 3\ \ 1\ \ 2\ \ 0\ \ 2\ \ 1\end{bmatrix}.
\end{equation}
For all of these rate options, the variable nodes in the derived graph that correspond to copies of the degree-$6$
variable node in the protograph (equivalently, the right-most column of the protomatrix) are punctured.

The name \emph{AR4JA} is derived from the operations reflected in the protograph structure and indicated in Figs.~\ref{fig_ar4ja12} and \ref{fig_ar4ja23}. As can be seen, the protograph embodies several features similar to those of an Accumulate-Repeat-Accumulate (ARA) code. For the AR4JA code construction, a partial precoding by accumulation (A) is followed by ``repetition $4$ times'' (R4), culminating in a ``jagged'' accumulation (JA). 
The jagged accumulation differs from a standard accumulation, which contains degree-two variable nodes only, by the additional edge at the upper right of the protograph.
In fact, the switch from a standard accumulation stage to the jagged accumulation stage, which reduces the number of degree-$2$ variable nodes, allows the AR4JA protographs to meet the criterion of the ensemble of protographs with linearly increasing minimum distance. 
More specifically, the techniques of Divsalar and other researchers \cite{DivDol06,DivDol09,AbuDiv11,Sweat} can be used to calculate the asymptotic ensemble weight enumerators for protograph-based codes, from which the \emph{typical relative minimum distance} $\delta_{\min}$ can be found.
They prove that the minimum distance $d_{\min}$ of most of the codes in the ensemble derived from the protograph exceeds $\delta_{\min} n$,
where $n$ is the block length of the code. %
For the rate-$1/2$ AR4JA protomatrix in (\ref{proto_ar4ja12}), $\delta_{\min} =0.015$ \cite{DivDol06}.

\section{QC Expansion and Polynomial Representation}
\label{sect3}
The codewords of a block code may be divided into non-overlapping \emph{subblocks} of $N$ consecutive symbols.
A \emph{quasi-cyclic} (QC) code is a linear block code having the property that applying identical circular shifts to every subblock of a codeword yields a codeword. 
QC codes are a generalization of conventional cyclic block codes and are simple to encode \cite[\textsection~8.14]{PetersonWeldon}.

A binary QC-LDPC code of length $n = L N$ can be described by an $m \times n$ sparse parity-check matrix $\matr{H}\in \F_2^{m\times n}$, with $m = J N$,
which is composed of $N \times N$ circulant submatrices.
A right \emph{circulant matrix} is a square matrix with each successive row right-shifted circularly one position relative to the row above.
Therefore, circulant matrices can be completely described by a single row or column. As in \cite{SmaVon12}, we use the description corresponding to the left-most column.

A binary QC-LDPC code can also be described in polynomial form, since there exists an isomorphism between the commutative ring of $N\times N$ circulant binary
matrices and the commutative ring of binary polynomials modulo $x^N-1$, \textit{i.e.}, $\F_2[x]/\langle{x^N \! - \! 1} \rangle$.
Addition and multiplication in the latter ring correspond to, respectively, addition and multiplication of polynomials in 
$\F_2[x]$, modulo ${x^N - 1}$.

The isomorphism between $N\times N$ binary circulant matrices and polynomial residues in the quotient ring
maps a matrix to the polynomial in which the coefficients in order of increasing degree correspond to 
the entries in the left-most matrix column taken from top to bottom.
Under this isomorphism, the $N\times N$ identity matrix maps to the multiplicative identity in the polynomial quotient ring, namely $1$. A few examples of the mapping (indicated by $\mapsto$) for $N = 3$ are shown below:

\begin{equation*}
\begin{bmatrix}
1\ 0\ 0\\
0\ 1\ 0\\
0\ 0\ 1\end{bmatrix} \mapsto 1 \quad
\begin{bmatrix}
0\ 0\ 1\\
1\ 0\ 0\\
0\ 1\ 0\end{bmatrix} \mapsto x \quad
\begin{bmatrix}
1\ 1\ 0\\
0\ 1\ 1\\
1\ 0\ 1\end{bmatrix} \mapsto 1+x^2.
\end{equation*}

This isomorphism requires that care be taken when representing multiplication of a circulant matrix $\matr{M}$ by a binary vector $\vect{v} = (v_0, v_1, \ldots, v_{N-1})$. 
If we associate the polynomial $M(x)$ with the matrix $\matr{M}$ under the isomorphism just described, and let $v(x) = v_0+v_1x+\cdots+ v_{N-1}x^{N-1}$ represent the vector $\vect{v}$, then 
the product $\left(\matr{M}\vect{v}^T\right)^T=\vect{v}\matr{M}^T$ maps to the polynomial $M(x) v(x)$ modulo $x^N-1$,
and the product $\vect{v}\matr{M}$ maps to the polynomial $x^N M(x^{-1}) v(x)$ modulo $x^N-1$.
Note that from this paragraph onwards, all indexing begins at zero and all vectors are row vectors.

Given a polynomial residue $a(x) \in \F_2[x]/\langle{x^N \!- \!1} \rangle$, we define its weight $\wt(a(x))\in\Z$ to be the number of nonzero coefficients.
Thus, the weight $\wt(a(x))$ of the polynomial $a(x)$ is equal to the Hamming weight $\wH(\vect{a})$ of the corresponding binary vector of coefficients $\vect{a}$.
For a length-$L$ vector of elements in the ring, $\vect{a}(x)=(a_0(x),a_1(x),\ldots,a_{L-1}(x))$, we define its Hamming weight to be the sum of the weights of its components, \textit{i.e.}, $\wH(\vect{a}(x))=\sum_{i=0}^{L-1} \wt(a_i(x))$.
Throughout this work, computations implicitly shift to integer arithmetic upon taking the weight.

In using a ring, there are a few important points to bear in mind. The elements in a ring $R$ need not have a multiplicative inverse; the ones that do are called \emph{units}.
The ring may include zero divisors, where a \emph{zero divisor} (or \emph{factor of zero}) is a nonzero element $a\in R$, such that $ab=0$ for some $b\in R$, $b\neq 0$. 
For example, the elements $a=2$ and $b=3$ in $R=\Z/6\Z$, the ring of integers modulo $6$, are zero divisors. 
Note that units cannot be zero divisors.
Moreover, in finite rings, such as $\Z/6\Z$ or $\F_2[x]/\langle{x^N \!-\! 1} \rangle$, 
every nonzero element of the ring must be either a unit or a zero divisor \cite[p.~205]{Pinter}.

The elements of weight one in the polynomial quotient ring $\F_2[x]/\langle{x^N \!-\! 1}\rangle$ are the monomials, all of which are units in the ring.
Specifically, the inverse of the monomial $a(x)=x^i$, $0\le i<N$, is the monomial $(a(x))^{-1}=x^{(N-i)\bmod N}$.
Under the isomorphism defined above, the monomials in the ring correspond to \emph{cyclic permutation matrices}, which are binary circulant matrices with a single one in each row and each column.

For any $N>0$, the nonzero elements in the ring $\F_2[x]/\langle{x^N \!-\! 1}\rangle$ that have even weight are zero divisors.
For instance, the product of the polynomial $x^{N-1}+x^{N-2}+\cdots+x+1$ and any even weight polynomial 
is zero. Odd weight polynomials, on the other hand, may sometimes be zero divisors, such as $x^3+x+1$ in the ring $\F_2[x]/\langle{x^7 \!-\! 1} \rangle$.

As we are interested in the connection between protographs and QC-LDPC codes, we focus on parity-check matrices $\matr{H}$ that are in $J \times L$ block matrix form, that is
\begin{equation*}
  {\matr{H}} \triangleq \left[ {\begin{array}{ccc}
     {{\matr{H}}_{0,0} } & \cdots & {{\matr{H}}_{0,L-1} }\\
      \vdots & \ddots & \vdots \\
     {{\matr{H}}_{J-1,0} } & \cdots & {{\matr{H}}_{J-1,L-1} }
   \end{array} } \right],
\end{equation*}
where each submatrix $\matr{H}_{j,i}$ is an $N \times N$ binary circulant matrix.
Let $h_{j,i,s} \in \F_2$ be the left-most entry in the $s$th row of the submatrix $\matr{H}_{j,i}$.
We can then write $\matr{H}_{j,i} = \sum_{s = 0}^{N-1} h_{j,i,s} \matr{I}_s$,
where $\matr{I}_s$ is the $N \times N$ identity matrix circularly left-shifted by $s$ positions.
Now, using the same convention as above for identifying matrices with polynomial residues, we can associate with
$\matr{H}$ the \emph{polynomial parity-check matrix} ${\matr{H}}(x)$, where
${\matr{H}}(x) \in \left(\F_2[x]/\langle{x^N \!-\! 1} \rangle \right)^{J \times L}$,
\begin{equation*}
 {\matr{H}}(x) \triangleq \left[ {\begin{array}{ccc}
    {h_{0,0} (x)} & \cdots & {h_{0,L-1} (x)}\\
     \vdots & \ddots & \vdots \\
    {h_{J-1,0} (x)} & \cdots & {h_{J-1,L-1} (x)}
  \end{array} } \right],
\end{equation*}
and $h_{j,i} (x) \triangleq \sum_{s = 0}^{N-1} {h_{j,i,s} } x^s$.

We will be interested in the weight of each polynomial entry of $\matr{H}(x)$, or,  equivalently, the row or column sum of each submatrix of $\matr{H}$.
The \emph{weight matrix} of $\matr{H}(x)$, which is a $J\times L$ matrix of nonnegative integers, is defined as
\begin{equation*}
\wt \left( {{\matr{H}}(x)} \right) \triangleq \left[ {\begin{array}{ccc} 
   {\wt \left( {h_{0,0} (x)} \right)} & \cdots & {\wt \left( {h_{0,L- 1} (x)} \right)} \\
    \vdots & \ddots & \vdots \\
   {\wt \left( {h_{J- 1,0} (x)} \right)} & \cdots & {\wt \left( {h_{J- 1,L- 1} (x)} \right)} \\
 \end{array} } \right].
\end{equation*}
Note that for a protograph-based QC-LDPC code, the weight matrix of the associated polynomial parity-check matrix
$\wt \left( {{\matr{H}}(x)} \right)$ is precisely the corresponding protomatrix, $\matr{A}=\wt \left( {{\matr{H}}(x)} \right)$.
It is also convenient to represent codewords of QC codes in polynomial form. In particular, the set of codewords, which is the set of vectors $\vect{c}$ such that ${\matr{H}}\vect{c}^T = \vect{0}^T$ over $\F_2$, maps to the set of polynomial vectors ${\vect{c}}(x) \in \left(\F_2[x]/\langle{x^N \!-\! 1} \rangle\right)^L$ satisfying ${\matr{H}}(x)\vect{c}(x)^T = \vect{0}^T$.
Under this identification, the entries $c_i(x)$ of the polynomial vector $\vect{c}(x) = \left( c_0(x), c_1(x), \ldots ,c_{L- 1}(x) \right)$ correspond to the length-$N$ subblocks of the codeword $\vect{c}$ that were 
defined at the start of this section.

\section{Minimum Distance Bounds for QC Codes}
\label{sect4}
In this section we review the upper bounds on the minimum (Hamming) distance of QC-LDPC codes that were established in \cite{SmaVon12} and then extend them to punctured QC-LDPC codes. 

\subsection{Upper Bounds for QC Codes}
\label{sect4.1}
We will use the shorthand notation $[L]$ to indicate the set of $L$ consecutive integers, $\left\{0,1,2,\ldots,L \!-\! 1 \right\}$. 
We also let $\set{S}\setminus i$ denote all the elements of $\set{S}$, excluding the element $i$.
We denote by $\matr{A}_\set{S}$ the submatrix of $\matr{A}$ containing the columns indicated by the index set $\set{S}$.
Similarly, $\vect{a}_{\set{S}}$ denotes the subvector containing the elements of the vector $\vect{a}$ indicated by the index set $\set{S}$.

The permanent of a $J \times J$ matrix $\matr{B} = \left[b_{j,i}\right]$ over a ring is defined to be
\begin{equation*}
\label{permdef}
\perm ({\matr{B}}) \triangleq \sum_\sigma {\prod_{j \in [J]} {b_{j,\sigma (j)}}},
\end{equation*}
where the summation is over all $J!$ permutations $\sigma$ of the set $[J]$, 
and $\sigma (j)$ is the $j$th entry of the permuted set $\sigma([J])$.
The definition of the permanent resembles that of the determinant of a square matrix, 
\begin{equation*}
\label{detdef}
\det ({\matr{B}}) \triangleq \sum_\sigma \sign(\sigma) {\prod_{j \in [J]} {b_{j,\sigma (j)}}},
\end{equation*}
where $\sign(\sigma)$ equals $+1$ if $\sigma$ is an even permutation and $-1$ if $\sigma$ is an odd permutation.
(Recall that an even permutation is obtained by applying an even number of transpositions of pairs of elements to the sequence $\{0,1,2,\ldots,J-1\}$.)
When the elements of ${\matr{B}}$ belong to a ring of characteristic two, where addition and subtraction are interchangeable, $\perm ({\matr{B}}) = \det ({\matr{B}})$. 
Like the determinant, the permanent may be computed recursively by taking the cofactor expansion along any row or column.
That is, for any $j \in [J]$, the cofactor expansion of the permanent of matrix $\matr{B}$ along the $j$th row is
\begin{equation}
\label{cofact}
\perm (\matr{B}) = \sum_{i \in [J]} b_{j,i} \cdot \perm (\matr{B}'_{[J]\setminus i}),
\end{equation}
with $\matr{B}'$ denoting the submatrix of $\matr{B}$ with the $j$th row removed. 
(Note that the subscript ${[J]\setminus i}$ on $\matr{B}'$ in (\ref{cofact}) removes the $i$th column.)

In the derivation of upper bounds on the minimum distance, we will make use of the notion of dependence
of vectors over a commutative ring $R$. This will require an appeal to some special properties of the ring of matrices over a finite commutative ring with unity. We remark that, in \cite{SmaVon12}, an alternative approach that exploits the connection between QC block codes and convolutional codes was used in the derivation of the upper bounds on the minimum distance.

Let $\set{S}$ be a set of vectors over $R$; that is, $\set{S}=\{\vect{s}_0,\ldots,\vect{s}_{n-1}\}$, where $\vect{s}_i \in R^N$.
The set $\set{S}$ is said to be \emph{dependent} if there exist elements $r_0,\ldots,r_{n-1}\in R$, not all zero,
such that the linear combination 
\begin{equation}
\label{lindep}
r_0 \vect{s}_0+\cdots+ r_{n-1} \vect{s}_{n-1}=(0,\ldots,0).
\end{equation}
If no such set of elements exists, the set $\set{S}$ is said to be \emph{independent}. 
(See, for example, \cite[p.~454]{Artin}.) 
We note that this independence test may be applied to the set of row vectors of a matrix with elements in $R$.
A \emph{dependent row} is any row in \eqref{lindep} which is multiplied by a nonzero scalar. 

\begin{lem}
\label{detdep}
Let $\matr{B}$ be a square matrix over a finite commutative ring with unity.
Then $\det (\matr{B})$ equals zero or is a zero divisor if and only if the set of row vectors of $\matr{B}$ is dependent.
\end{lem}
\begin{IEEEproof}
The proof of Lemma~\ref{detdep} can be found in the Appendix, along with some illustrative examples.
\end{IEEEproof}

We now review a technique from \cite{SmaVon12} for explicitly constructing codewords of a QC code specified by 
a polynomial parity-check matrix.

\begin{lem}[Lemma 6 \cite{SmaVon12}]
\label{lem1}
Let $\code{C}$ be a QC code with polynomial parity-check matrix 
$\matr{H}(x) \in \left(\F_2[x]/\langle{x^N \!-\! 1} \rangle\right)^{J \times L}$.
Let $\set{S}$ be an arbitrary size-$(J+1)$ subset of $[L]$ and let
$\vect{c}(x) \in \left(\F_2[x]/\langle{x^N \!-\! 1} \rangle\right)^L$
be a length-$L$ vector whose elements are given by
\begin{equation*}
c_i (x) \triangleq \begin{cases}
    {\perm \left( \matr{H}_{\set{S}\setminus i} (x) \right)}& \text{if } i\in \set{S} \\
    0 & \text{otherwise.}
\end{cases}
\end{equation*}
Then $\vect{c}(x)$ is a codeword in $\code{C}$.
\end{lem}
\begin{IEEEproof}
For any $j \in [J]$, let the $j$th row of $\matr{H}(x)$ be $\vect{h}_j (x)$. Then, 
\begin{align}
\notag
  \vect{h}_j (x) \vect{c}(x)^T 
  &= \sum_{i \in [L]} {h_{j,i} (x) \cdot c_i (x)} \\
\label{cwgen1a}
  &= \sum_{i \in \set{S}} {h_{j,i} (x) \perm \left( {{\matr{H}}_{\set{S}\setminus i} (x)} \right)} \\
\label{cwgen1b}
  &= \perm {\left[ {\begin{array}{*{20}c} 
	{{\vect{h}}_{j,\set{S}} (x)} \\
	{{\matr{H}}_{\set{S}} (x)}
\end{array} } \right]} \\
\label{cwgen1c}
  &= \det  {\left[ {\begin{array}{*{20}c}
	{{\vect{h}}_{j,\set{S}} (x)} \\
	{{\matr{H}}_{\set{S}} (x)}
\end{array} } \right]} = 0, 
\end{align}
where computations are in the ring $\F_2[x]/\langle{x^N \!-\! 1} \rangle$.
The cofactor expansion of \eqref{cwgen1b} is \eqref{cwgen1a}.
As the elements belong to a ring of characteristic two, the permanent in \eqref{cwgen1b} equals the determinant in \eqref{cwgen1c}. 
The determinant shown must be zero as it contains a repeated row. 
Since every row of ${\matr{H}}(x)$ has zero inner product with $\vect{c}(x)$, we conclude that 
${\matr{H}}(x) {\vect{c}}(x)^T = \vect{0}^T$. Therefore, ${\vect{c}}(x)$ is a codeword in $\code{C}$.
\end{IEEEproof}

While the $\min$ function applied to a collection of nonnegative real numbers returns the minimum, we will require a variant of this function, denoted $\minstarnoarg$, defined as follows. 
For a finite collection of nonnegative real numbers $\set{R}$, let $\set{R}^+ \subset \set{R}$ be the subset of positive elements of $\set{R}$.
We define
\begin{equation*}
\label{minst}
\minstarnoarg{\set{R}} \triangleq \begin{cases}
    \min{\set{R}^+} & \text{if } \set{R}^+ \neq \emptyset\\
    +\infty & \text{if } \set{R}^+ = \emptyset.
\end{cases}
\end{equation*}
The minimum distance of the QC code $\code{C}$ can then be written as 
\begin{equation}
\label{dmin}
d_{\min} (\code{C}) = \minstar{\vect{c}(x) \in \code{C}} {\wH \left( \vect{c}(x) \right)},
\end{equation}
where we have used $\minstarnoarg$ to exclude the all-zero codeword.

We now develop two possible upper bounds on the minimum distance. 
The first, based upon \cite{SmaVon12}, uses Lemma~\ref{lem1} to produce low-weight codewords from the polynomial parity-check matrix of the code. 
We will generate as many codewords as possible and apply (\ref{dmin}) to achieve an upper bound on the minimum distance. 
 
\begin{thm}[Theorem 7 \cite{SmaVon12}]
\label{thm0}
Let $\code{C}$ be a QC code with the polynomial parity-check matrix 
$\matr{H}(x) \in \left(\F_2[x]/\langle{x^N \!-\! 1}\rangle\right)^{J \times L}$.
Then the minimum distance of $\code{C}$ satisfies the upper bound
\begin{equation}
\label{distH0}
d_{\min} (\code{C}) \leq \minstar{\stacktwo{\set{S}\subseteq [L]}{|\set{S}| =J+ 1}}
  \sum_{i \in \set{S}}
  {\wt \left( {\perm \left( {{\matr{H}}_{\set{S}\setminus i} (x)} \right)} \right)}.
\end{equation}
\end{thm}
\begin{IEEEproof}
Let $\set{S}$ be a subset of $[L]$ of size-$(J+1)$ and apply Lemma \ref{lem1} to construct a codeword $\vect{c}(x)$ in $\code{C}$.
The weight of $\vect{c}(x)$ is
\begin{equation}
\label{distH0a}
\begin{split}
  \wH \left( {{\vect{c}}(x)} \right) 
  &= \sum_{i \in [L]} {\wt \left( {c_i (x)} \right)}\\
  &= \sum_{i \in \set{S}} {\wt \left( {\perm \left( {{\matr{H}}_{\set{S}\setminus i} (x)} \right)} \right)}.
\end{split}
\end{equation}
The upper bound (\ref{distH0}) follows immediately by combining (\ref{dmin}) and (\ref{distH0a}) and noting that, in general, only a strict subset of codewords can be generated by Lemma \ref{lem1}.
We must use the $\minstarnoarg$ function because, for some choices of the set $\set{S}$, the construction in Lemma \ref{lem1} will yield the all-zero codeword, and we have to exclude those sets from 
the calculation of the upper bound. 
\end{IEEEproof}

The second upper bound on the minimum distance makes use of an upper bound on the weight of the permanent of a matrix over the ring $\F_2[x]/\langle{x^N \!-\! 1}\rangle$, as described in the following lemma.

\begin{lem}
\label{lemtri}
Let $\matr{B}(x)$ be a $J \times J$ matrix with elements $b_{j,i}(x)$ in the ring $\F_2[x]/\langle{x^N \!-\! 1}\rangle$.
Then the weight of the permanent of $\matr{B}$ satisfies the upper bound
\begin{equation*}
\wt \left( \perm (\matr{B}(x)) \right) \leq
\perm \left( \wt(\matr{B}(x)) \right).
\end{equation*}
\end{lem}
\begin{IEEEproof}
Let the polynomials $a(x)$ and $b(x)$ be in the ring $\F_2[x]/\langle{x^N \!-\! 1}\rangle$.
We know that $\wt[a(x)+b(x)] \le \wt(a(x))+\wt(b(x))$,
as the maximum number of nonzero coefficients in the sum $a(x)+b(x)$ is $\wt(a(x))+\wt(b(x))$.

Similarly, we know that $\wt[a(x)\cdot b(x)] \le \wt(a(x)) \cdot \wt(b(x))$,
as the maximum number of nonzero coefficients in the product $a(x) \cdot b(x)$ is $\wt(a(x))\cdot\wt(b(x))$.
Therefore, 
\begin{equation*}
\label{lemtri2}
\begin{split}
\wt \left( \perm (\matr{B}(x)) \right) 
  &= \wt \bigg[\sum_\sigma {\prod_{j \in [J]} {b_{j,\sigma (j)}(x)}} \bigg]\\
  &\le \sum_\sigma \wt \left[ {\prod_{j \in [J]} {b_{j,\sigma (j)}(x)}} \right]\\
  &\le \sum_\sigma {\prod_{j \in [J]} \wt \left( {b_{j,\sigma (j)}(x)} \right)}\\
  &= \perm \left( \wt(\matr{B}(x)) \right).
\end{split}
\end{equation*}
\end{IEEEproof}

The second upper bound on the minimum distance, described in the next theorem, is expressed in terms of the weight matrix of 
the code. 

\begin{thm}[Theorem 8 \cite{SmaVon12}]
\label{thm00}
Let $\code{C}$ be a QC code with polynomial parity-check matrix 
$\matr{H}(x) \in \left(\F_2[x]/\langle{x^N \!-\! 1}\rangle\right)^{J \times L}$ and let $\matr{A}\triangleq \wt(\matr{H}(x))$.
Then the minimum distance of $\code{C}$ satisfies the upper bound
\begin{equation}
\label{distH00}
d_{\min} (\code{C}) \leq \minstar{\stacktwo{\set{S}\subseteq [L]}{|\set{S}| =J+ 1 }}
  \sum_{i \in \set{S}}
  {{\perm \left( {{\matr{A}}_{\set{S}\setminus i}} \right)}}
\quad\quad(\text{in } \Z).
\end{equation}
\end{thm}
\begin{IEEEproof}
The proof follows from Theorem~\ref{thm0} and Lemma \ref{lemtri}. 
The only subtlety arises from the fact that the sets $\set{S}$ excluded from (\ref{distH0}) by the $\minstarnoarg$ function may not be excluded from (\ref{distH00}) by the application of $\minstarnoarg$, \textit{i.e.}, there might be sets $\set{S}$ such that 
$\sum_{i \in \set{S}} \wt \left( \perm \left( {{\matr{H}}_{\set{S}\setminus i}(x)} \right)\right)=0$ but
$\sum_{i \in \set{S}} \perm \left( {{\matr{A}}_{\set{S}\setminus i}} \right)>0$.
The resolution of this potential complication can be achieved by reference to Theorem~8 in \cite{SmaVon12}. 
\end{IEEEproof}
\begin{remk}
Alternately, we may prove the correctness of this upper bound by following the arguments presented in the proof of Theorem~\ref{thm2}, below.
Since Theorem~\ref{thm2} includes puncturing, consider the set $\set{P}$ to be empty for this case.
\end{remk}

\subsection{Upper Bounds for Punctured QC Codes}
\label{sect4.2}
We now extend the preceding upper bounds on the minimum distance to the class of punctured QC codes.
The puncturing strategy of the AR4JA codes is prompted by the addition of the precoder to the underlying protographs, a modification that generally improves the decoding threshold \cite{DivDol06,DivDol09}.
Note that puncturing whole subblocks of the polynomial codeword ${\vect{c}}(x)$ preserves quasi-cyclicity. 
The set $\set{P}\subset [L]$ indexes the subblocks of the polynomial codeword ${\vect{c}}(x)$ which are not transmitted.
Indices of $\set{P}$ may also be associated with columns of the $J\times L$ polynomial parity-check matrix ${\matr{H}}(x)$.

We begin with a QC code $\code{C}$ based upon ${\matr{H}}(x)$.
Next, we define a new QC code $\code{C}'$ by puncturing the components  of ${\vect{c}}(x)$ that are indexed by $\set{P}$.
We mark the subblocks to be punctured with the symbol ``$\varphi$'' as re-indexing would introduce unnecessary notational complexity,
and we define $\wt(\varphi)= 0$, since the punctured symbols are not transmitted.

\begin{lem}
\label{lem2}
Let $\code{C}'$ be a punctured QC code constructed by puncturing subblocks of the QC code $\code{C}$,
defined by the polynomial parity-check matrix
${\matr{H}}(x) \in \left(\F_2[x]/\langle{x^N \!-\! 1}\rangle\right)^{J \times L}$.
Let the subblocks of ${\code{C}}$ indexed by the set $\set{P}$, $\set{P}\subset [L]$, be punctured.
Let $\set{S}$ be an arbitrary size-$(J+1)$ subset of $[L]$. 
Let the length-$L$ vector ${\vect{c'}}(x) = \left( {c_0' (x),c_1' (x),\ldots,c_{L- 1}' (x)} \right)$, with
$c'(x) \in \F_2[x]/\langle{x^N \!-\! 1}\rangle \cup \{ \varphi \}$, be defined by
\begin{equation*}
\label{cwgen2}
c_i' (x) \triangleq \begin{cases}
    {\perm \left( \matr{H}_{\set{S}\setminus i} (x) \right)}& \text{if } i\in \set{S}\setminus\set{P}\\
    \varphi\text{}& \text{if } i\in \set{P}\\
    0 & \text{otherwise}.\end{cases}
\end{equation*}
Then ${\vect{c'}}(x)$ is a codeword of the punctured code $\code{C'}$.
\end{lem}
\begin{IEEEproof}
This follows by noting that $\vect{c'}(x)$ is obtained by puncturing the subblocks indexed by $\set{P}$ 
from the codeword $\vect{c}(x)$ of Lemma \ref{lem1}.
\end{IEEEproof}

\begin{thm}
\label{thm1}
Let $\code{C'}$ be a punctured QC code constructed by puncturing subblocks of the QC code $\code{C}$ with polynomial 
parity-check matrix ${\matr{H}}(x) \in \left(\F_2[x]/\langle{x^N \!-\! 1}\rangle\right)^{J \times L}$.
Let the subblocks of ${\code{C}}$ indexed by the set $\set{P}$, $\set{P}\subset [L]$, be punctured.
Then 
\begin{equation}
\label{distH1}
d_{\min} (\code{C'}) \leq \minstar{\stacktwo{\set{S}\subseteq [L]}{|\set{S}| =J+ 1 }}
  \sum_{i \in \set{S}\setminus \set{P}}
  {\wt \left( {\perm \left( {{\matr{H}}_{\set{S}\setminus i} (x)} \right)} \right)}.
\end{equation}
\end{thm}
\begin{IEEEproof}
Let $\set{S}$ be a subset of $[L]$ of size-$(J+1)$, and apply Lemma \ref{lem2} to construct a codeword $\vect{c'}(x)$ in $\code{C'}$.
The weight of this codeword is
\begin{equation}
\label{distH1a}
\begin{split}
  \wH \left( {{\vect{c'}}(x)} \right) 
  &= \sum_{i \in [L]} {\wt \left( {c_i' (x)} \right)} \\
  &= \sum_{i \in \set{S}\setminus \set{P}} {\wt \left( {\perm \left( {{\matr{H}}_{\set{S}\setminus i} (x)} \right)} \right)},
\end{split}
\end{equation}
where we use the fact that $\wt(\varphi)=0$.
Combining (\ref{distH1a}) with (\ref{dmin}), we obtain the upper bound (\ref{distH1}), again noting that we obtain an upper bound because in general only a strict subset of codewords can be generated by Lemma \ref{lem2}.
\end{IEEEproof}

Care must be taken to ensure that the puncturing operation does not reduce the dimensionality of the code, that is, the base-2 logarithm of the number of distinct codewords. This is a requirement in the results that follow. 
Clearly, if puncturing a nonzero codeword of $\code{C}$ produces the all-zero codeword of $\code{C'}$,
dimensionality will be lost with respect to the original code.

\begin{lem}
\label{lem3}
Let $\code{C'}$ be a punctured QC code constructed by puncturing subblocks of the QC code $\code{C}$, while maintaining
the dimensionality of $\code{C}$.
Let the length-$L$ vector ${\vect{c}}(x)$ be a codeword of $\code{C}$ and ${\vect{c'}}(x)$ be a codeword of $\code{C'}$ obtained by puncturing ${\vect{c}}(x)$.
Then, ${c'_i}(x) \in \{0,\varphi\} \;\forall \, i \in [L]$ if and only if ${\vect{c}}(x) = \vect{0}$.
\end{lem}
\begin{IEEEproof}
The necessity of the condition ${\vect{c}}(x) = \vect{0}$ follows from the requirement that the dimensionality of the original code be maintained.
The sufficiency of the condition follows directly from the fact that puncturing the all-zero codeword of $\code{C}$ 
produces the all-zero codeword of $\code{C'}$. 
\end{IEEEproof}

Since the contribution of each subblock to the weight of the codeword is nonnegative, 
the weight of any particular punctured codeword must be less than or equal to its weight before puncturing, as can be seen by comparison of (\ref{distH1a}) to  (\ref{distH0a}).
Moreover, Lemma~\ref{lem3} implies that the upper bound of Theorem~\ref{thm1} will always be 
less than or equal to the upper bound of Theorem~\ref{thm0}, where no puncturing is used.

\begin{thm}
\label{thm2}
Let $\code{C'}$ be a punctured QC code constructed by puncturing subblocks of the QC code $\code{C}$, defined by the polynomial
parity-check matrix ${\matr{H}}(x) \in \left(\F_2[x]/\langle{x^N \!-\! 1}\rangle\right)^{J \times L}$
and let ${\matr{A}} \triangleq \wt \left( {{\matr{H}}(x)} \right)$.
Let the subblocks of ${\code{C}}$ indexed by the set $\set{P}$, $\set{P}\subset [L]$, be punctured, while maintaining the dimensionality of the code.
Then
\begin{equation}
\label{distA1}
d_{\min} (\code{C'}) \leq \minstar{\stacktwo{\set{S} \subseteq [L]}{ |\set{S}| = J + 1}}
  \sum_{i \in \set{S}\setminus \set{P}}
  {\perm \left( {{\matr{A}}_{\set{S}\setminus i} } \right)}
\quad\quad(\text{in } \Z).
\end{equation}
\end{thm}
\begin{IEEEproof}
The proof techniques we use are similar to those used in the proof of Theorem~8 in \cite{SmaVon12}, while avoiding 
the use of intermediate bounds based on convolutional codes.
Let $\set{S}$ be a subset of $[L]$ of size-$(J+1)$, and apply Lemma \ref{lem2} to construct a codeword, $\vect{c'}(x)$, in code $\code{C'}$.
From (\ref{distH1a}) we obtain
\begin{equation*}
\label{distA1a}
\begin{split}
  \wH \left( {{\vect{c'}}(x)} \right) 
  &= \sum_{i \in \set{S}\setminus \set{P}} {\wt \left( {\perm \left( {{\matr{H}}_{\set{S}\setminus i} (x)} \right)} \right)} \\
  & \leq \sum_{i \in \set{S}\setminus \set{P}} {\perm \left( {\wt \left( {{\matr{H}_{S\setminus i} }(x)} \right)} \right)} \\
  &= \sum_{i \in \set{S}\setminus \set{P}} {\perm \left( {{\matr{A}}_{\set{S}\setminus i} } \right)},
\end{split}
\end{equation*}
where we invoked Lemma \ref{lemtri} in the second step.

We now show the validity of using the $\minstarnoarg$ function in the upper bound.
The potential complication arises from the fact that for specific choices of the set $\set{S}$, namely those which yield the all-zero codeword in Lemma \ref{lem2},
the $\minstarnoarg$ function may not exclude their contribution to (\ref{distA1}), even though it does so in the bound (\ref{distH1}).
So, assume that a specific choice of $\set{S}$ yields $\wH(\vect{c'}(x))={0}$ according to (\ref{distH1a}), but produces a nonzero value for 
$\sum_{i \in \set{S}\setminus \set{P}} {\perm \left( {{\matr{A}}_{\set{S}\setminus i} } \right)}$. 
We must show there exists a nonzero codeword $\vect{c}^{*}{'}(x)$ for which 
$\wH (\vect{c}^{*}{'}(x))\le\sum_{i \in \set{S}\setminus \set{P}} {\perm \left( {{\matr{A}}_{\set{S}\setminus i} } \right)}$. 
Thus, for such specific choices of $\set{S} \subseteq [L]$, we assume in the remainder of the proof that ${c'_i}(x) \in \{0,\varphi\} \;\forall \, i \in [L]$. 
By Lemma \ref{lem3}, we know that ${\vect{c}}(x) =\vect{0}$ for this $\set{S}$. 
By Lemma \ref{lem1}, we know that every $J \times J$ submatrix of ${\matr{H}}_{\set{S}} (x)$ must have a zero permanent and determinant.
We now consider two cases.
\paragraph*{Case 1}
If $\sum_{i \in \set{S}\setminus \set{P}} {\perm \left( {{\matr{A}}_{\set{S}\setminus i} } \right)} = 0$,
then this specific $\set{S}$ has no effect on the bound (\ref{distA1}), as the zero result will be discarded by the $\minstarnoarg$ function.

\paragraph*{Case 2}
Alternatively, if $\sum_{i \in \set{S}\setminus \set{P}} {\perm \left( {{\matr{A}}_{\set{S}\setminus i} } \right)} > 0$,
there are no all-zero rows in $\matr{A}_{\set{S}}$ and, therefore, none in $\matr{H}_{\set{S}}(x)$.
However, we know that every $J \times J$ submatrix of ${\matr{H}}_{\set{S}} (x)$ has a zero determinant for
the specific $\set{S}$ that generates ${\vect{c}}(x) = \vect{0}$.
By Lemma~\ref{detdep}, this implies that the set of rows of $\matr{H}_{\set{S}}(x)$ is dependent.
We analyze this case further by setting aside the $t$th row of ${\matr{H}}_{\set{S}} (x)$, ${\vect{h}}_{t,\set{S}} (x)$,
preferring a dependent row to be the $t$th row\footnote{The proof holds no matter which row is chosen for removal; 
however, the row removal process terminates more rapidly if a dependent row is chosen.}.
We form a new matrix, ${\matr{H'}}(x)$, with the remaining $J-1$ rows of ${\matr{H}}(x)$ 
and a matrix, ${\matr{A'}}$, with the corresponding $J-1$ rows of ${\matr{A}}$. 
Because of the assumption that $\sum\nolimits_{i \in \set{S}\setminus \set{P}} {\perm \left( {{\matr{A}}_{\set{S}\setminus i} } \right)} > 0$,
there must be at least one index $i \in \set{S}\setminus \set{P}$, such that $\perm \left( {{\matr{A}}_{\set{S}\setminus i} } \right) > 0$.
The cofactor expansion of this term along row $t$ contains a term
\begin{equation}
\label{distA1b}
a_{t,i^*} \cdot \perm \left( {{\matr{A}}'_{(\set{S}\setminus i)\setminus i^{*}} } \right) > 0,
\end{equation}
for some $i^* \in \set{S}\setminus i$, where the positive integer $a_{t,i^*}$ is the entry in the $t$th row and $i^*$th column of $\matr{A}$.
Let $\set{S^{*}} \triangleq \set{S}\setminus i^{*}$.

Proceeding, we now assume that ${\matr{H'}}_{\set{S^{*}}}(x)$
contains at least one $(J-1) \times (J-1)$ submatrix with nonzero permanent. 
(If this is not true, we repeat the row removal process above, which may need to be repeated several times.
In the extreme case, these reductions could continue until we get a $1 \times 2$ matrix ${\matr{H'}}_{\set{S^{*}}} (x)$, having at least one nonzero entry.)
Then applying Lemma \ref{lem1}, with $\matr{H}(x)=\matr{H'}(x)$ and $\set{S}=\set{S^*}$, we generate a nonzero vector, $\vect{c}^{*}(x)$, with components
\begin{equation*}
\label{distA1c}
c_i^{*}(x) = \begin{cases}
   \perm \left( {{\matr{H}}'_{\set{S^*}\setminus i} (x)} \right) & \text{if } i \in \set{S^{*}}\\
   0 & \text{otherwise.}\end{cases}
\end{equation*}
The proof of Lemma \ref{lem1} implies that ${\matr{H'}}(x) \vect{c}^{*}(x)^T = \vect{0}^T$.
Multiplying the removed row of the parity-check matrix by the vector $\vect{c}^{*}(x)$ yields
\begin{equation*}
\label{distA1d}
\begin{split}
  {\vect{h}}_{t}(x) \vect{c}^{*}(x)^T
  &= \sum_{i \in [L]} {h_{t,i} (x) \cdot c_i^{*}(x)}\\
  &= \sum_{i \in \set{S^{*}}} {h_{t,i} (x) \perm \left( {{\matr{H}}'_{\set{S^{*}} \setminus i} (x)} \right)}\\
  &= \perm \left( {{\matr{H}}_{\set{S^{*}}} (x)} \right) = 0,
\end{split}
\end{equation*}
since all $J \times J$ submatrices of ${\matr{H}}_{\set{S}}(x)$ were assumed to have a zero permanent.
Therefore, the nonzero vector ${\vect{c}}^{*}(x)$ is a codeword in $\code{C}$. 

By puncturing $\vect{c}^{*}(x)$ we generate another nonzero vector, ${\vect{c}}^{*}{'}(x)$, which is a codeword in $\code{C'}$.
The Hamming weight of this codeword satisfies the upper bound 
\begin{align}
\notag
  \wH ({\vect{c}}^{*}{'}(x))
  &= \sum_{i \in \set{S}^*\setminus \set{P}} {\wt \left( {\perm \left( {{\matr{H}}'_{\set{S}^*\setminus i} (x)} \right)} \right)}\\
\label{distA1e2}
  &\leq \sum_{i \in \set{S^{*}}\setminus P} {\perm \left( {{\matr{A}}'_{\set{S}^{*}\setminus i} } \right)}\\
\label{distA1e3}
  &\leq a_{t,i^*} \cdot \sum_{i\in \set{S^{*}}\setminus P} {\perm \left( {{\matr{A}}'_{\set{S}^*\setminus i} } \right)}\\
\label{distA1e4}
  &\leq \sum_{i\in \set{S^{*}}\setminus \set{P}} \sum_{j\in \set{S}\setminus i} a_{t,j} \cdot \perm \left( {{\matr{A}}'_{(\set{S}\setminus i) \setminus j} } \right)\\
\label{distA1e5}
  &= \sum_{i\in \set{S^{*}}\setminus \set{P}} \perm \left( {{\matr{A}}_{\set{S}\setminus i} } \right)\\
\notag
  &\leq \sum_{i\in \set{S}\setminus \set{P}} \perm \left( {{\matr{A}}_{\set{S}\setminus i} } \right).
\end{align}
Applying Lemma \ref{lemtri} produces \eqref{distA1e2}.
Using the fact that $a_{t,i^*} \geq 1$, a consequence of (\ref{distA1b}), we upper bound \eqref{distA1e2} by \eqref{distA1e3}.
Next, we further upper bound \eqref{distA1e3} by \eqref{distA1e4} by adding additional nonnegative terms to \eqref{distA1e4}.
We recognize that \eqref{distA1e4} contains the sum of the cofactor expansions of each addend of \eqref{distA1e5}.

We can now conclude that, even if a set $\set{S}$ generates the all-zero codeword in Lemma \ref{lem2}, it
still yields a valid upper bound on the minimum distance of the punctured code, \textit{i.e.},
\begin{equation*}
\label{distA1f}
d_{\min}(\code{C'}) \leq \sum_{i \in \set{S}\setminus \set{P}} {\perm \left( {{\matr{A}}_{ \set{S}\setminus i} } \right)},
\end{equation*}
provided that $\perm(\matr{A}_{\set{S} \setminus i})$ is positive for at least one $i \in \set{S} \setminus \set{P}$.
The validity of the upper bound in (\ref{distA1}) follows.
\end{IEEEproof}

\section{Tighter Bounds on the Minimum Distance}
\label{sect5}
Examining the AR4JA protomatrices for rate-$2/3$ in (\ref{proto_ar4ja23}) and rate-$4/5$ in (\ref{proto_ar4ja45}), we see cases where the selection of 
$J+1=4$ columns of the weight matrix ${\matr{A}}$ will produce a submatrix ${\matr{A}}_\set{S}$ containing an all-zero top row.
This particular selection of $\set{S}$ produces the all-zero codeword by the codeword construction of Lemmas \ref{lem1} and \ref{lem2},
and, thus, will have no effect on the upper bounds of Theorems~\ref{thm0}, \ref{thm00}, \ref{thm1}, and \ref{thm2}.
We can improve those bounds by finding nonzero codewords after row elimination, as in the proof of Theorem~\ref{thm2}.

In the interest of brevity, we will state the following theorems in a way that applies to both unpunctured and punctured codes.
In the unpunctured case, it is understood that the set $\set{P}$ is empty.

\begin{lem}
\label{lem4}
Let ${\code{C}'}$ be a QC code constructed by optionally puncturing subblocks of the QC code ${\code{C}}$, defined by the 
polynomial parity-check matrix $\matr{H}(x)\in \left({\F_2[x]}/\langle{x^N \!-\! 1}\rangle\right)^{J\times L}$. 
Let the subblocks of ${\code{C}}$ indexed by the set $\set{P}$, $\set{P}\subset [L]$, be punctured.
Let $\matr{{H}'}(x)$ be a submatrix of $\matr{H}(x)$ with rows ${{\vect{h}}_{t}}(x)$, $t\in \set{T}\subset [J]$, removed. 
Let ${\set{S}}$ be a subset of $[L]$ of size $J+1-|\set{T}|$, such that
\begin{equation}
\label{tightcond}
\perm  {\left[ {\begin{array}{*{20}c}
   {{\vect{h}}_{t,\set{S}} (x)} \\
   {{\matr{H}}'_{\set{S}} (x)}
 \end{array} } \right]} =0 \;\forall\, t \in \set{T}.
\end{equation}
Let the components of the length-$L$ vector $\vect{{c}'}(x)$ = $\left( c'_0(x),c'_{1}(x),\ldots,c'_{L-1}(x) \right)$, with 
$c_i'(x)\in {\F_2[x]}/\langle{x^N \!-\! 1}\rangle \cup \{\varphi\}$, be defined as
\begin{equation*}
\label{tightcwgen1}
c_i'(x) = \begin{cases}
   \perm \left( {{\matr{H}}'_{\set{S}\setminus i} (x)} \right)&{\text{if } i \in \set{S}\setminus\set{P}} \\
   \varphi &{\text{if } i \in \set{P}} \\
   0&\text{otherwise.}
 \end{cases}
\end{equation*}
Then ${\vect{c'}}(x)$ is a codeword in $\code{C'}$.
\end{lem}
\begin{IEEEproof}
We consider two cases.
\paragraph*{Case 1}
If $\set{P}=\emptyset$ (the code is unpunctured), then $\code{C}=\code{C}'$.
We first examine every retained row of ${\matr{H}}(x)$, denoted by ${\matr{h}}_j (x)$, where $j \in [J]$ and $j \notin \set{T}$.
The inner product of ${\matr{h}}_j (x)$ with the vector $\vect{c'}(x)$ is
\begin{equation*}
\label{tightcwgen1a}
\begin{split}
  {\vect{h}}_j (x){\vect{c'}}(x)^T 
  &= \sum_{i \in \set{S}} {h_{j,i} (x) \perm \left( {{\matr{H}}'_{\set{S}\setminus i} (x)} \right)}\\
  &= \perm  {\left[ {\begin{array}{*{20}c}
   {{\vect{h}}_{j,\set{S}} (x)}\\
   {{\matr{H}}'_{\set{S}} (x)}
 \end{array} } \right]}  \hfill \\
  &= \det  {\left[ {\begin{array}{*{20}c}
   {{\vect{h}}_{j,\set{S}} (x)}\\
   {{\matr{H}}'_{\set{S}} (x)}
 \end{array} } \right]}  = 0, 
\end{split}
\end{equation*}
since the determinant expression contains a repeated row.
Next, for every row ${\matr{h}}_t (x)$ removed from ${\matr{H}}(x)$, \textit{i.e.}, every row ${\matr{h}}_t (x)$ in which $t \in \set{T}$, we have
\begin{equation*}
\label{tightcwgen1b}
\begin{split}
  {\vect{h}}_{t} (x) {\vect{c'}}(x)^T
   &= \perm  {\left[ {\begin{array}{*{20}c}
   {{\vect{h}}_{t,\set{S}} (x)} \\
   {{\matr{H}}'_{\set{S}} (x)} \end{array} } \right]} = 0, 
\end{split}
\end{equation*}
because the permanent was assumed to be zero in (\ref{tightcond}).
Since all rows of the original polynomial parity-check matrix ${\matr{H}}(x)$ have been accounted for,
${\matr{H}}(x)\vect{c'}(x)^T=\vect{0}^T$ and $\vect{c'}(x)$ is a codeword in  $\code{C}=\code{C'}$.

\paragraph*{Case 2}
If the code $\code{C'}$ is punctured, 
let the components of the length-$L$ vector $\vect{{c}}(x)=\left( c_0(x),c_{1}(x),\ldots,c_{L-1}(x) \right)$, with 
$c_i(x)\in {\F_2[x]}/\langle{x^N \!-\! 1}\rangle$, be defined as
\begin{equation*}
\label{tightcwgen1c}
c_i(x) = \begin{cases}
   \perm \left( {{\matr{H}}'_{\set{S}\setminus i} (x)} \right)&{\text{if } i \in \set{S}} \\
   0&\text{otherwise.}
 \end{cases}
\end{equation*}
The proof follows by noting that $\vect{c'}(x)$ is obtained by puncturing subblocks indexed by $\set{P}$ 
from the unpunctured codeword $\vect{c}(x)$, above.  Since Case 1 establishes that $\vect{c}(x) \in \code{C}$, we conclude that $\vect{c'}(x) \in \code{C'}$.
\end{IEEEproof}

Not only does Lemma \ref{lem4} remove all-zero rows from $\matr{H}_{\set{S}}(x)$, it also helps produce lower weight codewords in 
more general conditions, as the following example shows.
\begin{exmp}
Consider the polynomial parity-check matrix 
\begin{equation*}
\label{tightex1}
{\matr{H}}(x) = \left[ {\begin{array}{cccc}
   0 & 0 & 0 & f_1(x)\\
   x^a & x^b & x^c& f_2(x)\\
   x^a & x^b & x^d& f_3(x)\end{array} } \right],
\end{equation*}
where $f_i(x), \; i=1, 2, 3$ are arbitrarily chosen polynomials.
Since $\perm\left( {{\matr{H}}_{\set{S}}(x)} \right) = 0$ with the column set $\set{S}=\{0,1,2\}$, as required by \eqref{tightcond}, %
we proceed with single row removal on $\matr{H}(x)$.  
Application of Lemma \ref{lem4} for all possible choices of $\set{T}$ with $|\set{T}|=1$ yields the codewords $\vect{c}(x)=\vect{0}$ and
\begin{equation*}
{\vect{c}}(x) = \left( {x^{b + d} + x^{b + c} ,x^{a + d} + x^{a + c},0,0} \right) \bmod  {x^N -1}.
\end{equation*}
However, with careful consideration, we see that Lemma \ref{lem4} will let us delete two sub-rows when the column set is
$\set{S}=\left\{ {0,1} \right\}$. This  produces the obvious codeword 
$\vect{c}(x) = \left( {x^b ,x^a ,0,0} \right)$,
when $\set{T} = \left\{ {0,1} \right\}$ or  $\set{T} = \left\{ {0,2} \right\}$.
\end{exmp}

\begin{thm}
\label{thm3}
\label{thm5}
Let ${\code{C}'}$ be a QC code constructed by optionally puncturing subblocks of the QC code ${\code{C}}$, defined by the polynomial parity-check matrix 
$\matr{H}(x)\in \left({\F_2[x]}/\langle{x^N \!-\! 1}\rangle\right)^{J\times L}$. 
Let the subblocks of ${\code{C}}$ indexed by the set $\set{P}$, $\set{P}\subset [L]$, be punctured.
Let $\matr{{H}'}(x)$ be a submatrix of $\matr{H}(x)$ with rows ${{\vect{h}}_{t}}(x)$, $t\in \set{T}\subset [J]$, removed. 
Let $\set{S}$ be a subset of $[L]$ of size $J+1-|\set{T}|$, such that \eqref{tightcond} holds.
Then 
\begin{equation}
\label{tightdistH1}
d_{\min}(\code{C'})   \leq \minstar{\set{S},\set{T}} 
{\sum_{i \in \set{S}\setminus\set{P}} \wt \left( {\perm \left( \matr{H}'_{\set{S}\setminus i} (x) \right)} \right)}.
\end{equation}
\end{thm}
\begin{IEEEproof}
The proof mirrors the proof of Theorem~\ref{thm1}, with the weight of the resulting codeword now given by
\begin{equation*}
\label{tightdistH2}
  \wH \left( {{\vect{c'}}(x)} \right) = \sum_{i \in \set{S}\setminus \set{P}} {\wt \left( {\perm \left({\matr{H}'_{\set{S}\setminus i} (x)} \right)} \right)}.
\end{equation*}
\end{IEEEproof}

Note that the minimization in (\ref{tightdistH1}) requires the removal of every possible set of rows $\set{T}$ and 
every set of retained columns $\set{S}$ for which $|\set{S}|+|\set{T}|=J+1$ and (\ref{tightcond}) holds.
In the case of single row removal ($|\set{T}|=1$), any row may be removed, as our requirement (\ref{tightcond}) degenerates to the condition
$\perm \left( {{\matr{H}}_{\set{S}} (x)} \right) = 0$, which is independent of the row selected for removal. 
For multiple row removal, the conditions are more complex to evaluate as each row in the set to be removed must be tested individually to verify that (\ref{tightcond}) holds.

For a specified expansion factor $N$, Lemma~\ref{lem4} and Theorem~\ref{thm3} impose certain conditions on the set $\set{T}$ that allow for the removal of rows from the polynomial parity-check matrix. 
However, these conditions cannot be directly translated into a form applicable to the nonnegative weight matrix $\matr{A}$, which is independent of $N$.
Therefore, the following theorem uses the stricter condition that the sub-row ${{\vect{a}}_{t,\set{S}}}$ is all-zero before removal.
\begin{thm}
\label{thm4}
\label{thm6}
Let ${\code{C}'}$ be a QC code constructed by optionally puncturing subblocks of the QC code ${\code{C}}$, defined by the polynomial parity-check matrix 
$\matr{H}(x)\in \left(\F_2[x]/\langle{x^N \!-\! 1}\rangle\right)^{J\times L}$ and let $\matr{A}\triangleq {\wt}\left( \matr{H}(x) \right)$. 
Let the subblocks of ${\code{C}}$ indexed by the set $\set{P}$, $\set{P}\subset [L]$, be punctured, while maintaining the dimensionality of the code.
Let $\matr{{A}'}$ be a submatrix of $\matr{A}$ with rows ${{\vect{a}}_{t}}$, $t\in \set{T}\subset [J]$, removed. 
Let $\set{S}$ be a subset of $[L]$ of size $J+1-|\set{T}|$, such that the sub-rows ${{\vect{a}}_{t,\set{S}}}=\vect{0}$
$\forall \, t\in \set{T}$. 
Then 
\begin{equation*}
\label{tightAub}
 {{d}_{\min }}({\code{C}}')\le 
  \minstar{\set{S},\set{T}}
  {\sum_{i\in {\set{S}}\setminus \set{P}}{\perm\left( {{\matr{A}}'_{\set{S}\setminus i}} \right)}}
\quad\quad(\text{in } \Z).
\end{equation*}
\end{thm}
\begin{IEEEproof}
Let $\matr{{H}'}(x)$ be the submatrix of $\matr{H}(x)$ with rows ${{\vect{h}}_{t}}(x)$, $t\in \set{T}$, removed. 
The $|\set{T}|$ sub-rows of the weight matrix $\matr{A}$ to be removed are all-zero 
(\textit{i.e.}, ${\matr{a}}_{t,\set{S}} =\vect{0} \;\forall\, t \in \set{T}$) if and only if 
the corresponding sub-rows of the polynomial parity-check matrix $\matr{H}(x)$ are all-zero (\textit{i.e.}, ${\matr{h}}_{t,\set{S}}(x) =\vect{0} \;\forall\, t \in \set{T}$).
The latter condition implies that (\ref{tightcond}) holds and we may apply Lemma \ref{lem4}
with this $\set{S}$ and $\set{T}$ and construct a codeword $\vect{c'}(x)$ in the code $\code{C}'$.
By Theorem~\ref{thm3}, the weight of $\vect{c'}(x)$ is
\begin{equation*}
\begin{split}
\label{tightAwt}
  \wH \left( {{\vect{c'}}(x)} \right) 
&= \sum_{i \in \set{S}\setminus \set{P}} {\wt \left( {\perm \left({\matr{H}'_{\set{S}\setminus i} (x)} \right)} \right)}\\
&\le \sum_{i\in {\set{S}}\setminus \set{P}}{\perm\left( {{\matr{A}}'_{\set{S}\setminus i}} \right)},
\end{split}
\end{equation*}
where Lemma \ref{lemtri} is applied to obtain the inequality.
Once again, the use of the $\minstarnoarg$ function in the bound must be validated by consideration of the all-zero codewords discarded in Theorem~\ref{thm3}.
The reasoning largely parallels that used in the proof of Theorem~\ref{thm2}, but with $|\set{T}|$ 
sub-rows of $\matr{H}_\set{S}(x)$ guaranteed to be all-zero. We omit the details. 
\end{IEEEproof}

\begin{exmp}
The benefits of Theorem~\ref{thm4} may be seen by considering the weight matrix given by
\begin{equation*}
\matr{A}= \begin{bmatrix}
   0 & 0 & 0 & 2\\
   0 & 0 & 0 & 2\\
   1 & 2 & 2 & 1\end{bmatrix}.
\end{equation*}
Treating the code as unpunctured, Theorem~\ref{thm2} produces a minimum distance upper bound of $+\infty$, 
since all $3 \times 3$ submatrices of $\matr{A}$ have a zero permanent.
Theorem~\ref{thm4} produces a much tighter bound of $3$, when $\matr{A}'_\set{S}=\begin{bmatrix}1&2\end{bmatrix}$.
\end{exmp}

\begin{exmp}
We now consider the weight matrix
\begin{equation*}
\matr{A}= \begin{bmatrix}
   0 & 0 & 3 & 0 & 3 \\
   1 & 1 & 0 & 1 & 3 \\
   1 & 2 & 0 & 2 & 1 \end{bmatrix}
\end{equation*}
which resembles the rate-$1/2$ AR4JA protomatrix in (\ref{proto_ar4ja12}). 
Treating the code as unpunctured, Theorem~\ref{thm2} produces a minimum distance upper bound of $30$, 
while Theorem~\ref{thm4} produces a substantially tighter upper bound of $10$. 
The reason for the difference is that the permanents of Theorem~\ref{thm2} produce large values with the top row of $\matr{A}$ present.
Theorem~\ref{thm4} will remove the top row of $\matr{A}$ when the chosen column set is $\set{S}=\{0,1,3\}$, yielding the tighter bound.
\end{exmp}

\section{QC Expansion of AR4JA}
\label{sect6}
A direct QC expansion of the AR4JA protographs shown in Figs.~\ref{fig_ar4ja12} and \ref{fig_ar4ja23} will generate a QC-LDPC code. 
Applying Theorems \ref{thm2} and \ref{thm6} to the AR4JA protomatrices (\ref{proto_ar4ja12}) -- (\ref{proto_ar4ja45}) 
yields an upper bound of 10 on the minimum distance for all code rates, independent of block length. 
A minimum distance of $10$ or less is rather small for the large block lengths desired, motivating the consideration of a more involved expansion procedure.

In fact, the construction of the AR4JA codes defined in  \cite{CCSDS11} makes use of a two-step expansion process. 
After a first QC expansion (``lifting'') by a factor of $4$, a larger weight matrix is obtained, as illustrated for rate-$1/2$ by the matrix 
\begin{equation}
\label{bigA}
\matr{A} = \left[\begin{IEEEeqnarraybox*}[\mysmallarraydecl][c]{,c/c/c/c?c/c/c/c?c/c/c/c?c/c/c/c?c/c/c/c,}
0&	0&	0&	0&	0&	0&	0&	0&	1&	0&	0&	0&	0&	0&	0&	0&	1&	0&	0&	1\\
0&	0&	0&	0&	0&	0&	0&	0&	0&	1&	0&	0&	0&	0&	0&	0&	1&	1&	0&	0\\
0&	0&	0&	0&	0&	0&	0&	0&	0&	0&	1&	0&	0&	0&	0&	0&	0&	1&	1&	0\\
0&	0&	0&	0&	0&	0&	0&	0&	0&	0&	0&	1&	0&	0&	0&	0&	0&	0&	1&	1\vspace{.038in}\\
1&	0&	0&	0&	1&	0&	0&	0&	0&	0&	0&	0&	1&	0&	0&	0&	0&	1&	1&	1\\
0&	1&	0&	0&	0&	1&	0&	0&	0&	0&	0&	0&	0&	1&	0&	0&	1&	0&	1&	1\\
0&	0&	1&	0&	0&	0&	1&	0&	0&	0&	0&	0&	0&	0&	1&	0&	1&	1&	0&	1\\
0&	0&	0&	1&	0&	0&	0&	1&	0&	0&	0&	0&	0&	0&	0&	1&	1&	1&	1&	0\vspace{.038in}\\
1&	0&	0&	0&	0&	0&	1&	1&	0&	0&	0&	0&	1&	1&	0&	0&	1&	0&	0&	0\\
0&	1&	0&	0&	1&	0&	0&	1&	0&	0&	0&	0&	0&	1&	1&	0&	0&	1&	0&	0\\
0&	0&	1&	0&	1&	1&	0&	0&	0&	0&	0&	0&	0&	0&	1&	1&	0&	0&	1&	0\\
0&	0&	0&	1&	0&	1&	1&	0&	0&	0&	0&	0&	1&	0&	0&	1&	0&	0&	0&	1%
\end{IEEEeqnarraybox*}\right].
\end{equation}
This is a so-called \emph{type-$1$} weight matrix---that is, it contains only ones and zeros---implying that the associated protograph does not have parallel edges \cite{SmaVon12}. 

According to the CCSDS standard, the weight matrices so obtained, such as (\ref{bigA}), are considered to be protomatrices themselves, that are then expanded in a second QC expansion to create QC-LDPC codes with three block lengths, corresponding to $k=1024$, $4096$, and $16\,384$ information bits. 
For example, quasi-cyclically expanding (\ref{bigA}) by a factor of $N=128$ and puncturing the last 4 columns of \eqref{bigA} yields the $(n,k)=(2048,1024)$ AR4JA code.
In this final expansion, the binary parity-check matrix $\matr{H}$ is constructed by replacing each $1$ entry in (\ref{bigA}) by a cyclic permutation matrix selected using a variation on the ACE algorithm. 
These codes are QC with a subblock size equal to the second step expansion factor (\textit{e.g.}, $N=128$).
In other words, the two-step process is not equivalent to any single-step QC expansion. 

\begin{table}[!t]
\renewcommand{\arraystretch}{1.25} 
\caption{Minimum Distance of AR4JA Protomatrices After First QC Expansion (Independent of Block Length)}
\label{table_md2}
\centering
\begin{tabular}{c||c|c}
\hline
\bfseries Code & \bfseries Upper Bounds by & \bfseries Num.\ of sets $\set{S}$ of\\
\bfseries Rate $r$ & {\bfseries Theorems \ref{thm2} and \ref{thm6}} & \bfseries size $J+1$ in $[L]$ \\
\hline\hline
$1/2$ & $66$ & $7.8\times10^4$\\
$2/3$ & $58$ & $3.7\times10^7$\\
$4/5$ & $56$\rlap{\textsuperscript{a}} & $5.2\times10^{10}$\\
\hline
\multicolumn{3}{l}{\scriptsize \textsuperscript{a}Computations are not exhaustive in sets $\set{S}$ due to complexity.}%
\end{tabular}
\end{table}

To compute length-independent minimum distance bounds for the AR4JA codes specified in the CCSDS standard using the techniques we have developed in this paper, the protomatrices such as (\ref{bigA}) should be used.
The resulting upper bounds, shown in Table \ref{table_md2}, range from $56$ to $66$.
Note that, prior to \cite{SmaVon12}, the tightest known upper bound on the minimum distance of QC-LDPC codes was $(J+1)!$ in the case when the protomatrices are all-ones \cite{MKseagate,FossQC}. 
(We find in \cite{SmaVon12} that $d_{\min}\le(J+1)!$ also holds for the more general case of type-1 protomatrices.)
For the protomatrix of (\ref{bigA}), where $J=12$, this would yield the extremely loose upper bound $d_{\min} \le 6.2 \times 10^9$.

This example points to the potential advantage of a two-step QC expansion, as suggested by the increase from $10$ to the range $56$ -- $66$ in the minimum distance upper bound. 
It also illustrates the strength of the general class of hierarchical QC-LDPC codes that have recently been examined in \cite{HierarchQC}.

In order to reduce the computation time required to produce the results in Table \ref{table_md2}, a number of techniques were used. 
For larger weight matrices, if we assume that calculations are dominated by the computation time $t_J$ for the $J \times J$ permanent,
then the total time to evaluate Theorem~\ref{thm2} is $t_J (J+1) \binom{L}{J+1}$.
The final term $\binom{L}{J+1}$ is the number of sets $\set{S}$ in the weight matrix $\matr{A}$ and is shown in the right column of Table \ref{table_md2} for AR4JA.
For the computations of interest, with $J=12$, we built a simple recursive routine with $t_{12}=44 \mu s$ for computing sparse permanents (as measured on a 2.6 GHz CPU).
Thus, the estimated time for computing the rate-$1/2$ results in Table \ref{table_md2} is $44s$, while we measured an actual run-time of $53s$, including bookkeeping and set manipulations.
On the other hand, for the rate-$4/5$ construction, the estimated time to completely evaluate the upper bound of Theorem~\ref{thm2} is $344$ days. 
Therefore, we selectively directed the computations, yielding the results shown in Table \ref{table_md2}. 
These selective computations were performed by choosing sets $\set{S}$ which, based upon the findings from the rate-$2/3$ code, we thought would yield the smallest value permanents in evaluating \eqref{distA1}.
Additional efforts at each code rate to recompute the minimum distance bounds using the row elimination logic of Theorem~\ref{thm4} did not yield tighter results with the AR4JA weight matrices.

\begin{table}[!t]
\renewcommand{\arraystretch}{1.25} 
\caption{Distance of CCSDS AR4JA Parity-Check Matrix}
\label{table_md3}
\centering
\begin{tabular}{c||c|c|c|c}
\hline
\bfseries Code & \multicolumn{2}{c|}{\bfseries Minimum Distance } & \multicolumn{2}{c}{\bfseries Stopping Distance }\\
\bfseries Rate & \multicolumn{2}{c|}{\bfseries U.B.\ by Searching} & \multicolumn{2}{c}{\bfseries U.B.\ by Searching }\\
\cline{2-5}
\bfseries $r$ & \bfseries $k=1024$ & \bfseries $k=4096$ & \bfseries $k=1024$ & \bfseries $k=4096$\\
\hline\hline
$1/2$ & $52$ & $63$ & $50$ & $63\rlap{\textsuperscript{a}}$\\
$2/3$ & $26$ & --  &     $23$ & $62\rlap{\textsuperscript{b}}$\\
$4/5$ & $13$ & $27$ & $11$ & $25$\\
\hline
\multicolumn{5}{l}{\scriptsize \textsuperscript{a}The smallest stopping set found was a codeword.}\\
\multicolumn{5}{l}{\scriptsize \textsuperscript{b}Beyond the upper bound shown in Table \ref{table_md2}.}%
\end{tabular}
\end{table}

\begin{table}[!t]
\renewcommand{\arraystretch}{1.25} 
\caption{Estimated Weight Spectrum for CCSDS AR4JA Rate-$4/5$, $k=1024$ using
Search Parameters: $I=150$ and $T=10$}
\label{table_s1}
\centering
\begin{tabular}{l||c|c|c|c||c}
\cline{2-6}
   & \multicolumn{4}{c||}{\bfseries Hamming Weight} & \bfseries Search \\
\cline{2-5}
   & $13$ & $14$ & $15$ & $16$ & \bfseries Time \\
\hline\hline
\bfseries Num.\ of & & & & & \\
\bfseries Codewords & $32$ & $256$ & $128$ & $400$ & 1.5 hrs\\
\hline
\end{tabular}
\end{table}

\begin{table}[!t]
\renewcommand{\arraystretch}{1.25} 
\caption{Estimated Weight Spectrum for CCSDS AR4JA Rate-$4/5$, $k=4096$ using
Search Parameters: $I=300$ and $T=19$}
\label{table_s2}
\centering
\begin{tabular}{l||c|c|c||c}
\cline{2-5}
  & \multicolumn{3}{c||}{\bfseries Hamming Weight} & \bfseries Search \\
\cline{2-4}
  & $27$ & $28$ & $29$ & \bfseries Time \\
\hline\hline
\bfseries Num.\ of & & & \\
\bfseries Codewords & $128$ & $0$ & $0$ & 82 hrs\\
\hline
\end{tabular}
\end{table}

\begin{table}[!t]
\renewcommand{\arraystretch}{1.25} 
\caption{Estimated Weight Spectrum for CCSDS AR4JA Rate-$2/3$, $k=1024$ using
Search Parameters: $I=300$ and $T=19$}
\label{table_s3}
\centering
\begin{tabular}{l||c|c|c|c||c}
\cline{2-6}
  & \multicolumn{4}{c||}{\bfseries Hamming Weight} & \bfseries Search \\
\cline{2-5}
  & $26$ & $29$ & $31$ & $32$ & \bfseries Time \\
\hline\hline
\bfseries Num.\ of & & & & & \\
\bfseries Codewords & $64$ & $128$ & $64$ & $64$ & 5 hrs\\
\hline
\end{tabular}
\end{table}

\section{Numerical Results Obtained by Search}
\label{sect_search}

In this section, we present numerical results on the minimum distance of AR4JA codes obtained by means of computer search for low-weight codewords. We also examine bounds on the girth of AR4JA codes.

\subsection{Distance Bounds from Codeword Search for AR4JA Codes}

Several papers, including \cite{Richter06,HuDist,DecFosImpulse}, have described search techniques to find the minimum distance and/or stopping distance of general LDPC codes. 
To validate our bounds on the minimum Hamming distance, we utilized the error impulse and decoding algorithm of \cite{Richter06} to conduct a non-exhaustive search for small stopping sets, and then examined the results to identify the codewords. 
We modified the algorithm to take advantage of QC symmetry by skipping impulse combinations which are identical after cyclically shifting every subblock.
We also broadened the search space by increasing the value of the parameters in \cite{Richter06} corresponding to the number of iterations $I$ and the maximum threshold $T$.
In addition, before the erasure decoding step, we erased the punctured symbols in addition to the symbols already erased by the algorithm.
The resulting upper bounds on the minimum distance and the stopping distance obtained by this search methodology are summarized in Table \ref{table_md3}.%

It should be noted that, in \cite{Richter06}, the search algorithm was applied to rate-$1/2$ codes up to a minimum distance of $19$, and the estimated weight spectrum results were listed only up to a maximum codeword weight of $25$. 
It may be the case, then, that the application of this method to some of the AR4JA codes under consideration here may be pushing the algorithm beyond its effective range. Thus, further searching may turn up lower weight codewords and stopping sets than we have found.

We note that, as a consequence of the quasi-cyclicity of the code, when a codeword of a QC-LDPC code is located by our search technique, it is an indication of a group of codewords with the same weight.
For instance, the low-weight codewords of the rate-$4/5$, $k=1024$ CCSDS AR4JA code generally occur as a set of $N=32$ cyclically-shifted versions of a base codeword. 
On occasion, when all subblocks are periodic with a common period, the cyclically-shifted codeword returns to the base codeword after only a fraction of $N$ shifts.
Accounting for this, we tabulated all distinct low-weight codewords found using our search algorithm for three of the AR4JA parity-check matrices given in the CCSDS standard. The estimated weight spectra, based upon the codewords we identified during our search, are shown in Tables \ref{table_s1} -- \ref{table_s3}.

\begin{figure}[!t]
\centering
\includegraphics[width=\figwidth]{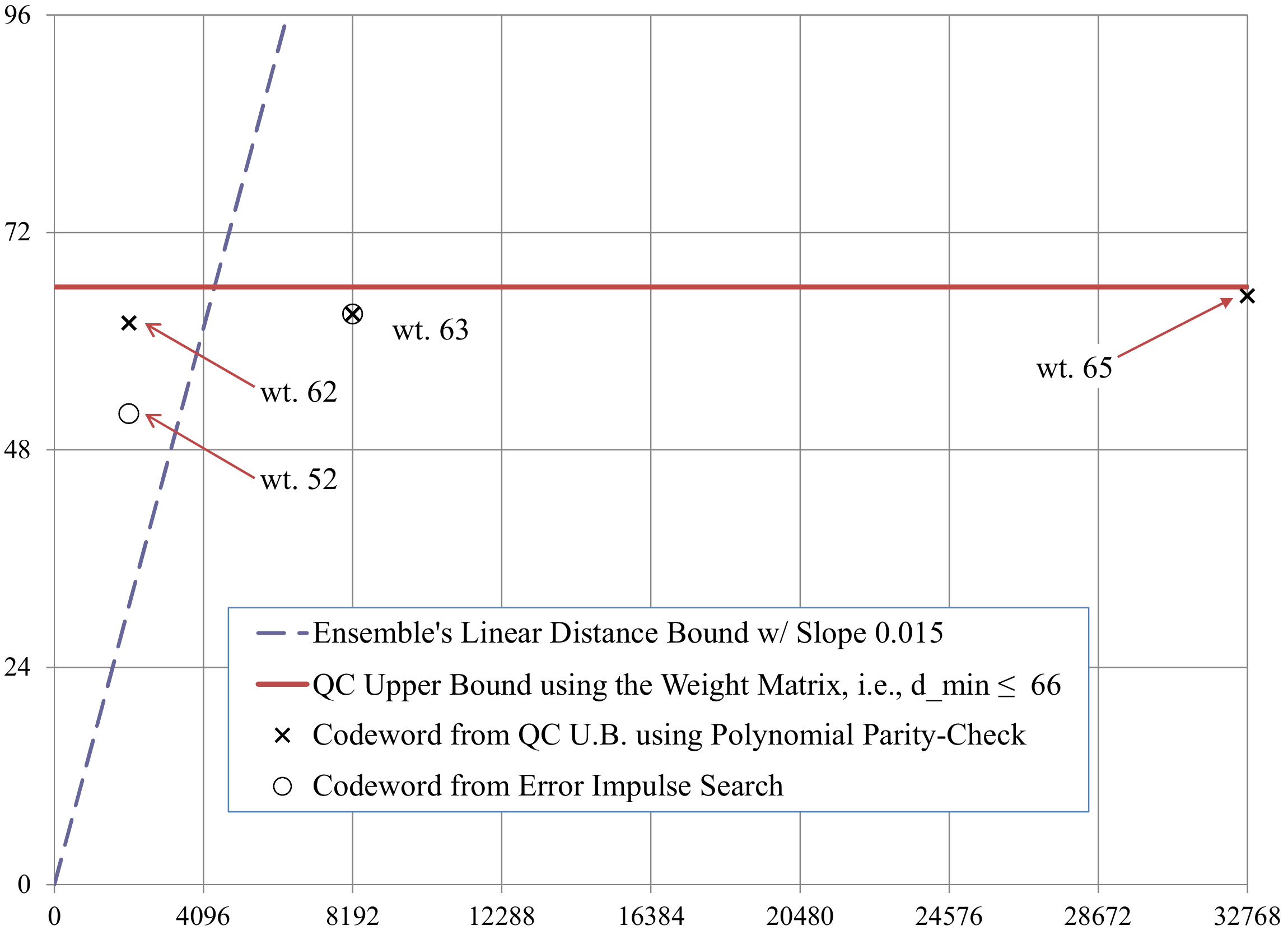}
\caption{Minimum distance bounds vs. block length for rate-$1/2$ AR4JA.}
\label{fig_distsum}
\end{figure}

Fig.~\ref{fig_distsum} summarizes our minimum distance results for rate-$1/2$ AR4JA codes as a function of the block length $n$.
First, the upper bound of $66$ obtained from the weight matrix using Theorems \ref{thm2} and \ref{thm4} is plotted as a horizontal line.
The points indicated by the symbol ${\times}$ represent the minimum distance upper bounds based on the QC polynomial parity-check matrices from Theorem~\ref{thm1}. These bounds are $62$, $63$, and $65$ for the block lengths $2048$, $8192$, and $32\,768$, respectively.
Finally, the codeword weights found by our search technique, as shown in Table \ref{table_md3}, are designated by
the symbol $\circ$.
Note that the smallest codeword weights found in our search are fairly close to the length-independent bound of $66$ for all block lengths.
We note that while our codeword searches generally used error impulse pairs, the codeword of weight $52$ at the smallest block length was found using impulse triplets.

Divsalar et al.\ showed that most codes in the ensemble of certain protograph-based codes, including the AR4JA codes,
have minimum distance linearly increasing with block length \cite{DivDol06,DivDol09,AbuDiv11}.
Specifically, by upper bounding the ensemble average weight enumerator, they were able to prove that
$\Pr \left\{ d_{\min} < \delta_{\min} n\right\} \to 0$ exponentially fast as $n \to \infty$, for some constant $\delta_{\min} >0$. 
For the rate-$1/2$ AR4JA-based ensemble of codes, they computed the value $\delta_{\min} = 0.015$. 
The corresponding linear growth of the ensemble minimum-distance is plotted in Fig.~\ref{fig_distsum} as a dashed line.  

It can be seen that for the QC-AR4JA codes that appear in the standard, our bounds are tighter than the linearly increasing ensemble bound for $n \geq 4400$ bits. By examining the probability that a random expansion is quasi-cyclic, we can shed some light on this possibly surprising situation.
Consider the expansion of each $1$ entry in a protomatrix such as (\ref{bigA}) by a factor $N$.
There are $N$ cyclic permutation matrices to choose from and $N!$ general permutation matrices.
Thus, a randomly chosen permutation matrix has only a probability of $1/(N \!-\! 1)!$ of being cyclic.
This probability goes to zero super-exponentially fast.
Since the QC class of expansions is such a small fraction of the ensemble of all possible expansions, 
one cannot claim with certainty that the probabilistic bound of Divsalar et al.\ applies to the resulting class of codes.

\subsection{Bounds on the Girth of AR4JA Codes}

In this section, we show that the two-step expansion approach was essential in order to achieve girths beyond $6$ for larger block lengths of the rate-$1/2$ AR4JA codes.
Recall that the \emph{girth} of a code denotes the length of the shortest cycle in its Tanner graph. 
Table~\ref{table_girth} summarizes the results of our calculations of the girth of the AR4JA codes.
The girths of the standardized codes for each block length and code rate are shown in the table. Also
shown, in parentheses, are upper bounds obtained by the tree method of \cite{Gal63,Sweat}, which we briefly
describe. In the protograph, the neighborhood of any node can be diagrammed as a tree.
We can measure how tall this tree is at a given number of nodes corresponding to the specified block length.
We do this for each node type in the protograph and select the smallest as an upper bound on girth that would apply to any possible expansion method. 
These are the upper bounds recorded in the table.

We also determined upper bounds on the girth that derive from properties of QC expansions.
Since the AR4JA protomatrices of (\ref{proto_ar4ja12}) -- (\ref{proto_ar4ja45}) all contain the element $3$,
the girth of the derived graphs obtained by QC expansion cannot exceed $6$ \cite{SmaVon12}.
However, since the codes in the CCSDS standard use a two-step expansion, we must examine the intermediate protomatrices such as (\ref{bigA}) to evaluate the girth.
We find that they contain 
the submatrix $\bigl[ \begin{smallmatrix} 1&1&1\\1&1&1\end{smallmatrix} \bigr]^T$ at every code rate.
This limits the girth of the QC expansion to a maximum of $12$, independent of block length \cite{ParkHongISIT,FossQC,Tan01}, as shown in the final row of Table \ref{table_girth}.

\begin{table}[!t]
\renewcommand{\arraystretch}{1.25} 
\caption{Girth of the CCSDS AR4JA codes}
\label{table_girth}
\centering
\begin{tabular}{r||c|c|c}
\hline
\bfseries Information & \multicolumn{3}{c}{\bfseries Measured Girth (Upper Bound)}\\
\cline{2-4}
\bfseries Bits $k$ & $r=1/2$ & $r=2/3$ & $r=4/5$\\
\hline\hline
$1024$  & $6\ (12)$ & $4\ (10)$ & $4\ (8)$\\
$4096$  & $8\ (14)$ & $6\ (10)$ & $4\ (10)$\\
$16384$ & $10\ (16)$ & $6\ (12)$ & $4\ (10)$\\
\hline 
\hline 
QC Limit of& & & \\
Protomatrix& $12$ & $12$ & $12$\\
\hline
\multicolumn{4}{l}{\scriptsize (\#) Denotes upper bound computed by tree method \cite{Gal63,Sweat}}%
\end{tabular}
\end{table}

\section{Conclusion}
\label{sect_concl}
This work has extended the upper bounds on the minimum distance of QC-LDPC codes developed in \cite{SmaVon12} to the class of punctured QC-LDPC codes. 
We have also tightened those distance bounds in situations where the codes are derived from protographs whose protomatrices contain many zeros.

We evaluated the minimum distance upper bounds for the AR4JA codes specified in the CCSDS standard for deep space communication. 
Our results show that the use of a two-step expansion in the definition of these codes was critical to achieve reasonably high minimum distance. 
On the other hand, we have also shown that the minimum distance of the standardized QC AR4JA codes does not grow with block length, even though the asymptotic ensemble minimum distance of AR4JA codes grows linearly in the block length \cite{DivDol06,DivDol09,AbuDiv11}.
Nevertheless, the minimum distance of the CCSDS codes is likely high enough for practical purposes. 

The bounds developed here and in \cite{SmaVon12} can be useful tools in evaluating future QC-LDPC code designs, both punctured and unpunctured.

\appendix[Proof of Lemma~\ref{detdep}]
In this appendix we state several properties of matrices over a commutative ring with unity.
Several terms defined in Sections \ref{sect3} and \ref{sect4} will be used here, such as determinant, unit, zero divisor, and independence.

\subsection{Matrices Over Commutative Rings}

Let $R$ be a commutative ring and let
$\mathcal{M}_N(R)$ be the ring of $N \times N$ matrices over $R$.
The determinant of a matrix $\matr{A}\in \mathcal{M}_N(R)$ is denoted by $\det (\matr{A})$. 
Given matrices $\matr{A}, \matr{B} \in \mathcal{M}_N(R)$, we have the identity \cite[\textsection~12.2]{Artin}
\begin{equation}
\label{detprob}
\det (\matr{A}\matr{B}) = \det(\matr{A}) \det(\matr{B}).
\end{equation}

If $R$ is a commutative ring with unit element $1$, then a matrix $\matr{A}$ is a unit in $\mathcal{M}_N(R)$ if and only if $\det(\matr{A})$ is a unit in $R$. 
To see this, suppose that $\matr{A}$ has inverse $\matr{A}^{-1}$. 
From (\ref{detprob}), we conclude that $\det(\matr{A}^{-1})\det (\matr{A})=\det (\matr{I})=1$, where $\matr{I}$ is the $N \times N$ identity matrix. 
This implies that $\det(\matr{A})$ is a unit in $R$. 
Conversely, from Cramer's rule, we have $\det(\matr{A}) \matr{I} = \matr{A} \adj(\matr{A})$, where $\adj(\matr{A})$ is the adjugate (or classical adjoint) 
of $\matr{A}$ \cite{Artin}. If $\det(\matr{A})$ is a unit in $R$, then $\matr{A}$ is invertible, with inverse 
$\matr{A}^{-1}=(\det(\matr{A}))^{-1} \adj(\matr{A})$.


Consider the free module of rank $N$ over $R$, that is, $R^N=\{\vect{r}=(r_0,\ldots,r_{N-1})| r_i \in R, \forall i=0,\ldots,N-1 \}$. 
The set $\mathcal{V}=\{\vect{v}_0,\ldots, \vect{v}_{n-1}\}$, where $\vect{v}_i \in R^N$, $i=0, \ldots, n-1$, \emph{generates} $R^N$ if every $\vect{r} \in R^N$ can be expressed as a linear combination of members of the set $\mathcal{V}$, \textit{i.e.}, $\vect{r}=\sum_{i=0}^{n-1} a_i \vect{v}_i $, with $a_i \in R$, for all $i=0, \ldots, n-1.$
Furthermore, a generating set $\mathcal{V}$ is a \emph{basis} for $R^N$ if it is also independent. 
The \emph{standard basis} $\{\vect{e}_0,\ldots,\vect{e}_{N-1}\}$ of $R^N$ is simply the set of rows of the $N \times N$ identity matrix $\matr{I}$.

The following theorem is a reformulation of material from \cite[ch.~5]{Connell}.

\begin{thm}
\label{thm_app1}
Let $R$ be a commutative ring with unity, and let $\matr{A} \in \mathcal{M}_N(R)$ have rows
$\vect{v}_0,\ldots,\vect{v}_{N-1} \in R^N$ and columns $\vect{w}_0,\ldots,\vect{w}_{N-1} \in R^N$.
The following statements are equivalent.
\begin{enumerate}[\IEEEsetlabelwidth{4)}]
\item $\matr{A}$ is invertible, \textit{i.e.}, a unit in $\mathcal{M}_N(R)$.
\item $\det \matr{A}$ is a unit in $R$.
\item The set $\{\vect{v}_0,\ldots,\vect{v}_{N-1}\}$ generates $R^N$.
\item The set $\{\vect{w}_0 ,\ldots,\vect{w}_{N-1}\}$ generates $R^N$.
\end{enumerate}
\end{thm}
\begin{IEEEproof}
We have already shown in the discussion above the equivalence between 1) and 2).
We now prove that 1) implies 3). 
Namely, if $\matr{A}$ is invertible, then for any $\vect{r} \in R^N$ we can write
$$
\vect{r}=\vect{r} \matr{I}=(\vect{r} \matr{A}^{-1}) \matr{A}.
$$
Conversely, if the rows of $\matr{A}$ generate $R^N$, then they must generate the elements of the standard basis. 
Given vectors $\{\vect{u}_0, \ldots, \vect{u}_{N-1}\} \in R^N$ satisfying $\vect{u}_i \matr{A} = \vect{e}_i$, for all $i=0, \ldots, N-1$, the inverse matrix $\matr{A}^{-1}$ is the matrix whose rows are the vectors $ \vect{u}_0, \ldots, \vect{u}_{N-1}$.

Finally, the equivalence of 3) and 4) follows from the fact that 
$\det (\matr{A})=\det{(\matr{A}^T)}$, which implies that $\det{(\matr{A}^T)}$ is a unit in $R$ if and only $\det{(\matr{A})}$ is.
\end{IEEEproof}

\subsection{Finite Commutative Rings}
In our application, we consider finite rings with unity.
As already noted, every nonzero element of a finite ring must be either a unit or a zero divisor.

\begin{thm}
\label{thm_app2}
Let $R$ be a finite commutative ring with unity and let $\matr{A} \in \mathcal{M}_N(R)$.
The rows $\vect{v}_0,\ldots,\vect{v}_{N-1}$ of $\matr{A}$ generate $R^N$ if and only if they are independent.
\end{thm}
\begin{IEEEproof}
If the rows of $\matr{A}$ generate $R^N$, then, by Theorem~\ref{thm_app1}, $\matr{A}$ is invertible and not a zero divisor in $\mathcal{M}_N(R)$.
Therefore, if a matrix $\matr{B} \in \mathcal{M}_N(R)$ satisfies $\matr{B}\matr{A}=\matr{0}$, then $\matr{B}=\matr{0}$, 
proving that the rows of $\matr{A}$ are independent.
Conversely, suppose the rows of $\matr{A}$ are independent. If they do not generate $R^N$, then,  by the pigeonhole principle, there exist distinct vectors $\vect{r},\vect{s} \in R^N$ such 
that $\vect{r} \matr{A}=\vect{s} \matr{A}$. 
This implies that $\vect{r} \matr{A} - \vect{s} \matr{A} = (\vect{r}-\vect{s}) \matr{A} = (0,\ldots,0)$. This contradicts the assumed independence of the rows of $\matr{A}$, so the rows of $\matr{A}$ must generate $R^N$. 
\end{IEEEproof}

Theorems~\ref{thm_app1} and \ref{thm_app2} imply that if $R$ is a finite commutative ring with unity and $\matr{A} \in \mathcal{M}_N(R)$, then the determinant $\det (\matr{A}) $ is zero or a zero divisor in $R$ if and only if the rows of $\matr{A}$ are dependent. 
Of course, the ring used throughout this work, $\F_2[x]/\left<x^N-1\right>$, is a finite commutative ring with unity.

\begin{exmp}
(a)
Consider the infinite ring of integers with $R=\Z$. Define the matrix
\begin{equation*}
\matr{A}= \begin{bmatrix}
   1 & 0 \\
   0 & 2 \end{bmatrix}.
\end{equation*}
Note that $\det (\matr{A})=2$, and that $2$ is not a unit in $\Z$.
By Theorem~\ref{thm_app1}, the matrix $\matr{A}$ is not invertible and, therefore, its rows cannot generate all of $\Z^2$. In particular, no linear combination of the rows of $\matr{A}$ can generate odd integers in the second coordinate. On the other hand, the rows are independent, as can be easily verified.

(b)
Next, consider $R=\Z/3\Z$, the finite ring of integers modulo-$3$. Here, the ring element $2$ is a unit, since $2\cdot2=1$. By Theorem~\ref{thm_app1}, $\matr{A}$ is invertible and, therefore, its rows generate $R^2$. In accordance with Theorem~\ref{thm_app2}, the rows are independent.

(c)
Finally, consider $R=\Z/6\Z$, the finite ring of integers modulo-$6$. In this case, $2$ is a zero divisor, so, by Theorem~\ref{thm_app1}, the rows of $\matr{A}$ do not generate $R^2$. 
By Theorem~\ref{thm_app2}, there is at least one dependent row. 
Specifically, multiplying the second row by the scalar $3\in R$ yields the zero element $(0,0) \in R^2$.
\end{exmp}

\begin{exmp}
(a)
Consider the infinite polynomial ring $R=\F_2[x]$. Define the square matrix 
\begin{equation*}
\matr{A}= \begin{bmatrix}
   1 & 1 \\
   0 & 1+x \end{bmatrix}.
\end{equation*}
Note that $ \det (\matr{A})=1+x,$ which is not a unit in $R$.
Therefore, by Theorem~\ref{thm_app1}, the matrix $\matr{A}$ is not invertible and its rows do not generate $R^2$. Yet, the rows are independent.

(b)
Consider the finite quotient ring $R=\F_2[x]/\langle{x^N \!-\! 1}\rangle$. In this ring, $1+x$ is a zero divisor. Therefore, $\matr{A}$ is not invertible and the rows do not generate $R^2$. Since $R$ is finite, Theorem~\ref{thm_app2} implies that the set of rows is dependent. 
Specifically, multiplying the second row by the scalar $x^{N-1}+x^{N-2}+\cdots+x+1\in R$ yields the zero element $(0,0) \in R^2$.
\end{exmp}
\section*{Acknowledgment}
The authors gratefully acknowledge contributions to the proofs by Pascal Vontobel and Lance Small.
The authors would also like to thank Dariush Divsalar for useful discussions.



\bibliographystyle{IEEEtran}
\bibliography{IEEEabrv,Butler}

\end{document}